\documentclass[conference]{IEEEtran}
\IEEEoverridecommandlockouts
\usepackage{cite}
\usepackage{amsmath,amssymb,amsfonts}
\usepackage{graphicx}
\usepackage{textcomp}
\usepackage{xcolor}

\usepackage{graphicx}
\usepackage{amsmath}
\usepackage{caption}
\usepackage{subcaption}
\usepackage{algpseudocode}
\usepackage{algorithm}
\usepackage{algorithmicx}
\usepackage{algpascal}
\usepackage{url}
\usepackage{footnote}
\usepackage[font=footnotesize]{caption}
\usepackage{commath}
\usepackage{dblfloatfix}
\usepackage{lipsum}
\usepackage{mwe}
\usepackage{color}
\usepackage{hyperref}

\urldef{\sparrowurl}\url{https://people.eecs.berkeley.edu/~matei/papers/2013/sosp_sparrow.pdf}

\makesavenoteenv{tabular}
\makesavenoteenv{table}
\def\BibTeX{{\rm B\kern-.05em{\sc i\kern-.025em b}\kern-.08em
    T\kern-.1667em\lower.7ex\hbox{E}\kern-.125emX}}
\begin{document}

\title{AntDT: A Self-Adaptive Distributed Training Framework for Leader and Straggler Nodes}

\author{\IEEEauthorblockN{Youshao Xiao,
Lin Ju,
Zhenglei Zhou,
Siyuan Li,
Zhaoxin Huan}
\IEEEauthorblockN{
Dalong Zhang,
Rujie Jiang,
Lin Wang, 
Xiaolu Zhang,
Lei Liang,
Jun Zhou\textsuperscript{*}
}
\IEEEauthorblockA{
\textit{Ant Group, Hangzhou, China}\\
\{youshao.xys, julin.jl, zhouzhenglei.zzl, lisiyuan.li, zhaoxin.hzx, dalong.zdl, rujie.jrj\}@antgroup.com \\
\{fred.wl, yueyin.zxl, leywar.liang, jun.zhoujun\}@antgroup.com
}
\thanks{
*Corresponding author}
}

\maketitle

\begin{abstract}
Many distributed training techniques like Parameter Server and AllReduce have been proposed to take advantage of the increasingly large data and rich features. However, stragglers frequently occur in distributed training due to resource contention and hardware heterogeneity, which significantly hampers the training efficiency. Previous works only address part of the stragglers and could not adaptively solve various stragglers in practice. Additionally, it is challenging to use a systematic framework to address all stragglers because different stragglers require diverse data allocation and fault-tolerance mechanisms. Therefore, this paper proposes a unified distributed training framework called AntDT (Ant Distributed Training Framework) to adaptively solve the straggler problems. Firstly, the framework consists of four components, including the \emph{Stateful Dynamic Data Sharding service}, \emph{Monitor}, \emph{Controller}, and \emph{Agent}. These components work collaboratively to efficiently distribute workloads and provide a range of pre-defined straggler mitigation methods with fault tolerance, thereby hiding messy details of data allocation and fault handling. Secondly, the framework provides a high degree of flexibility, allowing for the customization of straggler mitigation solutions based on the specific circumstances of the cluster. Leveraging this flexibility, we introduce two straggler mitigation solutions, namely AntDT-ND for non-dedicated clusters and AntDT-DD for dedicated clusters, as practical examples to resolve various types of stragglers at Ant Group. Justified by our comprehensive experiments and industrial deployment statistics, AntDT outperforms other SOTA methods more than 3$\times$ in terms of training efficiency. \textcolor{black}{Additionally, in Alipay's homepage recommendation scenario, using AntDT reduces the training duration of the ranking model from 27.8 hours to just 5.4 hours.} 


\end{abstract}


\section{Introduction}\label{sec:introduction}
With the great success of deep learning on increasingly large data, distributed training techniques draw more attention to speed up the training procedure in the industry ~\cite{xin2021production,verbraeken2020survey,xiao2023adaptive,li2014scaling}. Among these techniques, data parallelism has emerged as one of the most popular methods for scaling out model training. The data parallelism approach involves partitioning the entire dataset evenly into mutually exclusive subsets at the beginning of training. Each worker then performs computations based on the assigned data partition and synchronizes the resulting local gradients globally in each iteration to update the model parameters iteratively~\cite{li13pytorch,rajbhandari2020zero}.

There are two main architectures designed for data-parallel distributed training in deep learning, namely the Parameter Server architecture~\cite{li2014scaling} and the AllReduce architecture~\cite{sergeev2018horovod}. In a typical Parameter Server (PS) architecture, there are servers and workers, with the servers responsible for storing, aggregating, and updating the model parameters. On the other hand, each worker pulls the latest parameters from servers, executes the main computation, and pushes back the intermediate computed results to servers for global synchronization in each iteration. \textcolor{black}{Depending on different consistency models~\cite{li2014scaling,ho2013more}, the Parameter Server is classified into Bulk Synchronous Parallel (BSP), Stale Synchronous Parallel (SSP), and Asynchronous Parallel (ASP) ~\cite{dean2012large,ho2013more,cui2014exploiting,xing2015petuum}. 
In BSP mode, the synchronization barrier enforces that each worker cannot begin the next iteration until all workers complete the computation and finish the global synchronization. ASP does not require synchronization, while SSP comes in a middle ground between two schemes by allowing the leading workers to proceed ahead of the stragglers with bounded iterations. In contrast, the AllReduce paradigm, commonly used in GPU clusters, involves only workers and synchronizes using the BSP consistency model~\cite{sergeev2018horovod,paszke2019pytorch,xiao2023g}.}

\begin{figure}
\begin{minipage}[t]{0.48\linewidth}
\includegraphics[width=\linewidth]{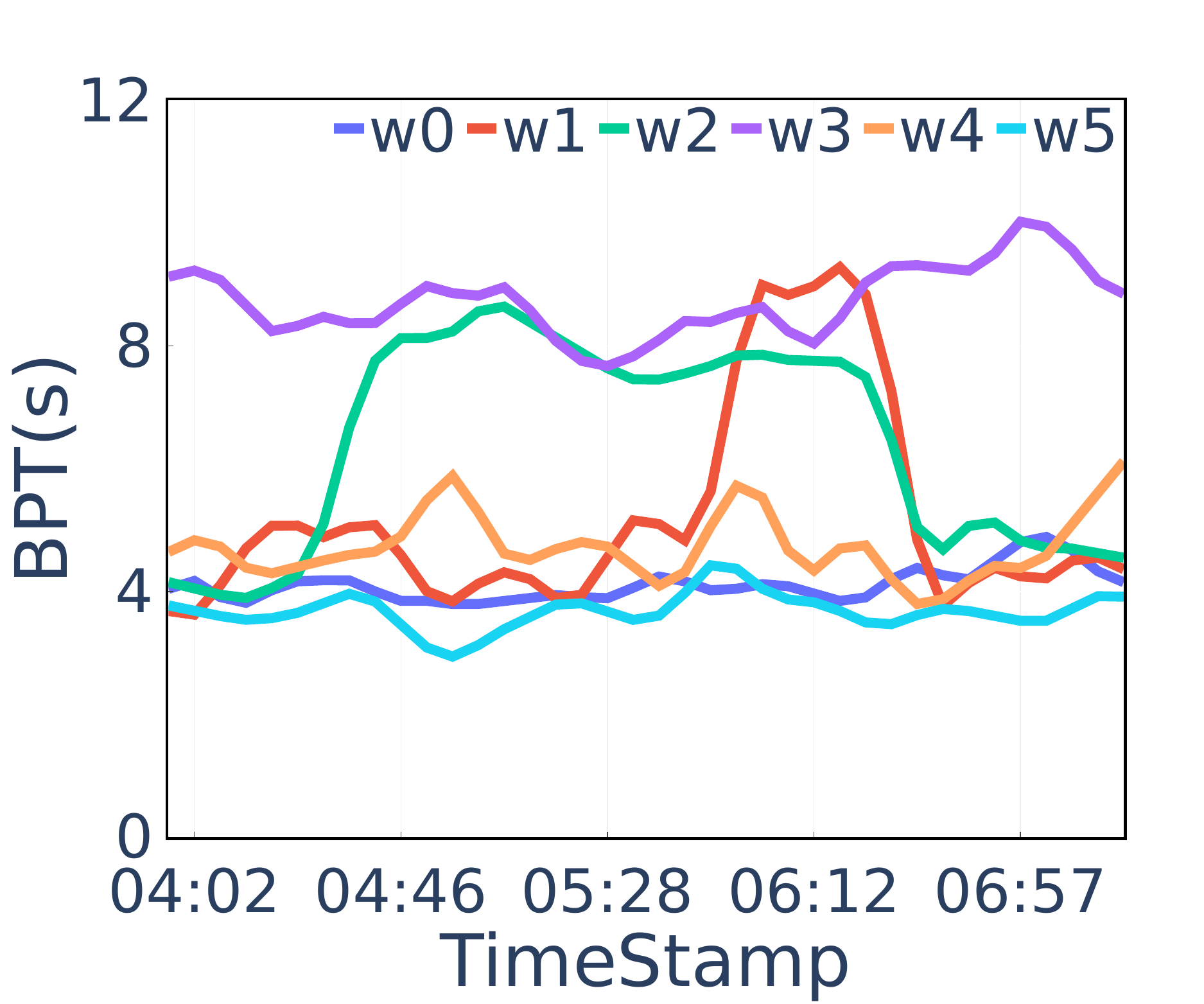} %
\subcaption{BPT(s) among six workers.}
\label{fig:obs-bpt-workers}
\end{minipage}
\hfill
\begin{minipage}[t]{0.48\linewidth}
\includegraphics[width=\linewidth]{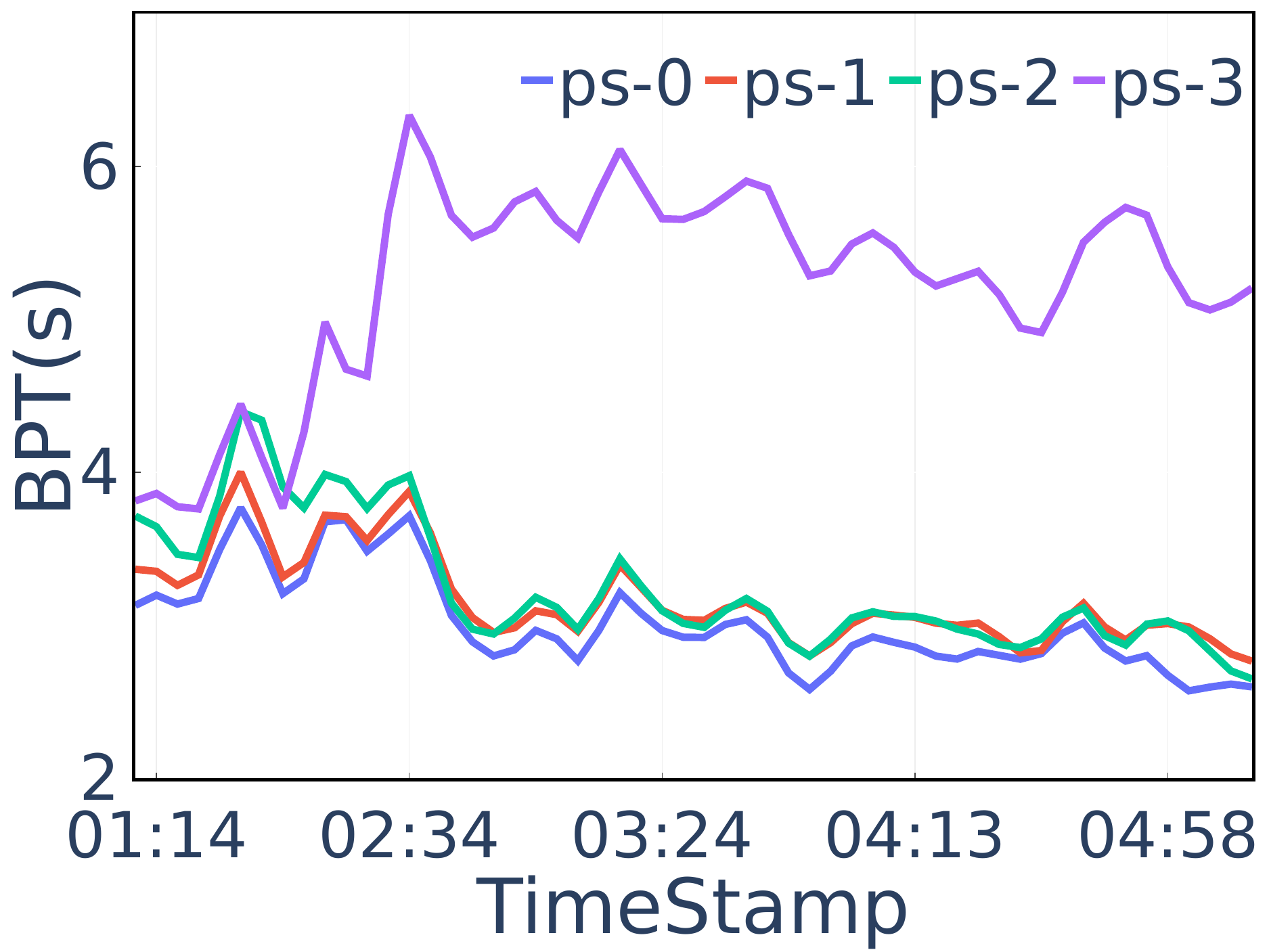} %
\subcaption{BPT(s) among four servers.}
\label{fig:obs-bpt-servers}
\end{minipage}
\caption{Batch Processing Time (BPT) among workers and servers in non-dedicated CPU cluster at Ant Group Cloud.}

\end{figure}

However, the efficiency of training can be significantly hampered by the straggler problem in the public cloud. \textcolor{black}{\textbf{Straggler nodes} \emph{refer to slow nodes that impede the progress of other normal nodes that also known as} \textbf{leader nodes}, \emph{until they finish their assigned workloads}\cite{ananthanarayanan2010reining,ananthanarayanan2013effective}. This issue arises due to two primary reasons, namely hardware heterogeneity and resource contention\cite{harlap2016addressing,gill2020tails}. Hardware heterogeneity or deterioration is the first reason, especially for the public cloud consisting of older and newer series of devices. We refer to these stragglers as \textbf{deterministic stragglers}, \textit{which are caused by deterministic performance gaps between different series of devices, leading to distinct training speeds}}. For instance, V100 GPUs are consistently about three times faster than P100 in GPU clusters. We also could find similar trends in CPU devices, as shown by the worker w3 in Fig. \ref{fig:obs-bpt-workers}.

\begin{figure}
\begin{minipage}[t]{0.5\linewidth}
    \includegraphics[width=\linewidth]{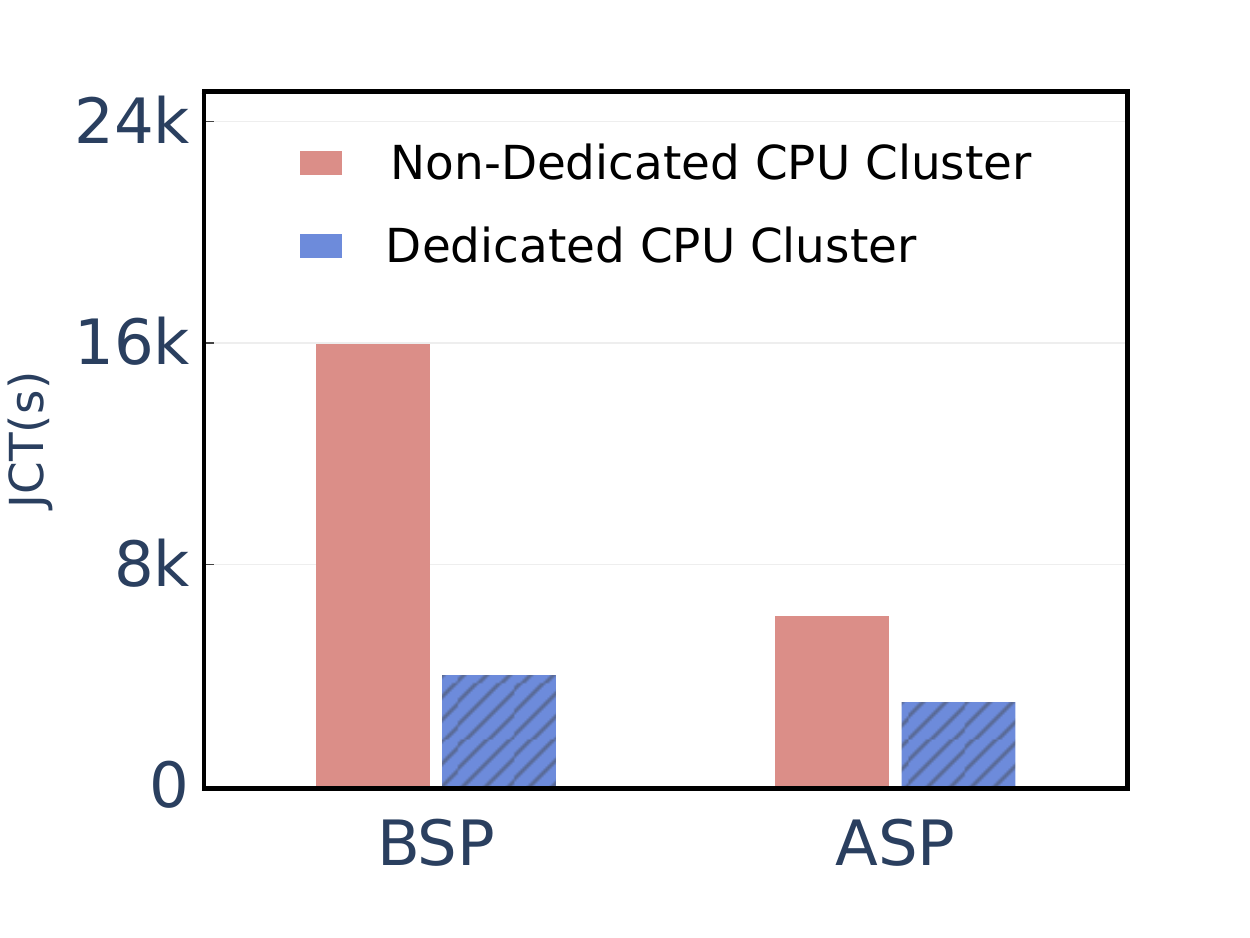}
    \caption{Job completion time (JCT) between BSP and ASP in dedicated and non-dedicated CPU clusters using XDeepFM~\cite{lian2018xdeepfm} model.}%
    \label{fig:obs-jct-bsp-asp}%
\end{minipage}%
\hfill
\begin{minipage}[t]{0.45\linewidth}
\includegraphics[width=\linewidth]{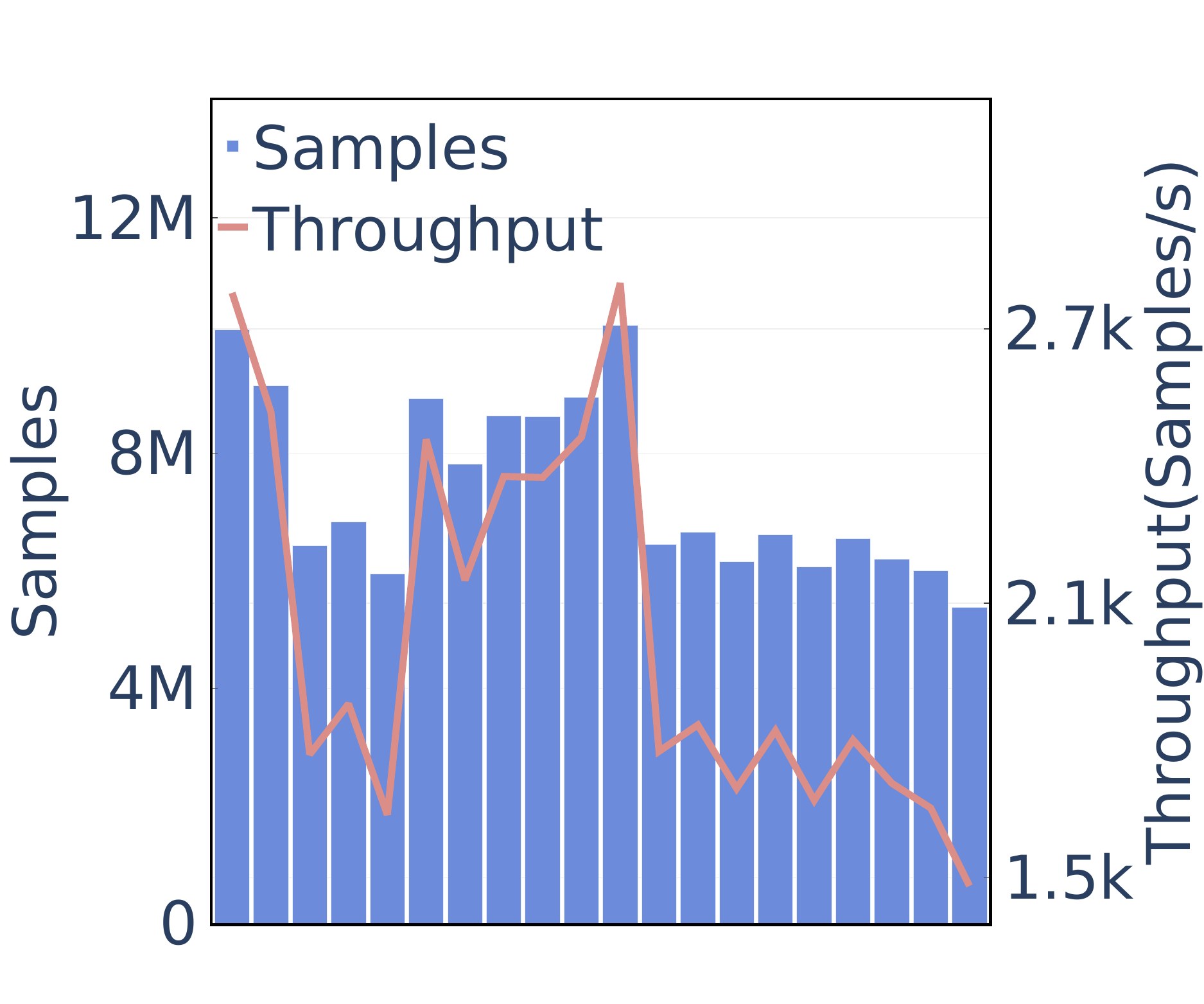}
\caption{Data consumption and local throughput among workers in ASP of Parameter Server in the non-dedicated CPU cluster.}
\label{fig:obs-data-consumption}
\end{minipage}
\end{figure}

Resource contention is another reason for straggler problems, which causes non-deterministic stragglers. \textcolor{black}{Non-dedicated clusters and dedicated clusters are two main clusters used in cloud vendors. \textbf{Non-dedicated clusters} \textit{refer to clusters shared by multiple tenants, e.g. spot instances}, where \textbf{dedicated clusters} \textit{composed of dedicated devices, which is exclusively occupied by one tenant}\cite{chen2020semi,burns2016borg,ec2,aliyun}}. \textcolor{black}{In non-dedicated clusters, heterogeneous workload scheduling~\cite{bernstein2014containers,harlap2017proteus,verma2015large,lu2020understanding} mainly leads to stragglers problems in batch jobs, such as machine learning training jobs.} Heterogeneous workload scheduling collocates latency-sensitive production jobs and training jobs in the same cluster with separate priorities~\cite{burns2016borg,lu2020understanding}, leading to stragglers when a training job is assigned to the same machine as production jobs. This problem is illustrated in Fig. \ref{fig:obs-jct-bsp-asp}, where the training speed in non-dedicated CPU clusters is on average four times slower than in dedicated CPU clusters for either BSP or ASP mode in the Ant Group cloud cluster. These stragglers are non-deterministic and we classify them into Transient Stragglers and Persistent Stragglers according to the periodicity of stragglers.\textcolor{black}{\textbf{Transient Stragglers} \textit{are stragglers with low throughput in the short term} illustrated by the worker w1 in Fig. \ref{fig:obs-bpt-workers}.  \textbf{Persistent Stragglers}  \textit{are consistently slower than other nodes over a long-term period caused by resource contention}, such as worker w2 in Fig.~\ref{fig:obs-bpt-workers} and ps-3 in Fig. \ref{fig:obs-bpt-servers}.}

\textcolor{black}{The distributed training community tries to address these issues from various aspects. Firstly, load-balancing-based methods~\cite{zhou2020falcon,harlap2016addressing,chen2020semi,tyagi2020taming} try to rebalance the workloads assigned to straggler and leader workers\footnote{We assume the parameters stored on the servers are evenly distributed.}. It usually has a low time cost and is effective for transient stragglers as well as deterministic stragglers. Secondly, replication-based methods\cite{chen2016revisiting, hanna2020adaptive,tandon2017gradient} usually launch duplicate tasks of identified stragglers
and only accept the results from the first finished tasks in a
job. These methods may lose a few samples in abandoned straggler nodes, potentially compromising the statistical performance of models. Furthermore, these methods require a data allocation mechanism that can efficiently distribute the samples to different workers due to the unbalanced consumption of samples among leader and straggler workers as depicted in Fig. \ref{fig:obs-data-consumption}. However, these approaches rely on customized and complex data assignment mechanisms, rendering them incompatible with one another. Additionally, both methods are ineffective for persistent stragglers caused by resource contention in both computing or network~\cite{chen2020elastic}.  Therefore, from the cluster scheduler perspective, scheduling-based methods~\cite{or2020resource,wu2021elastic,peng2018optimus,harlap2017proteus} try to solve the problem by killing the lagging node, relaunching a new node, and resuming the training by recovering from the periodically saved training states, known as checkpoints~\cite{abadi2016tensorflow,paszke2019pytorch}. However, this method is time-consuming, since it includes the time of scheduling and resuming the training states. Hence it incurs high time costs for transient stragglers. }

In this paper, we argue that existing stragglers approaches can only solve certain specific types of stragglers, lacking a unified framework that can adaptively handle various types of stragglers in industrial scenarios. However, addressing all stragglers in a systematic framework poses serval challenges due to the inherent diversity in their requirements for data allocation and fault-tolerance mechanisms. Therefore, we propose AntDT, a self-adaptive distributed training
framework for leader and straggler nodes in the data-parallel distributed training. Firstly, AntDT framework comprises four key components: the \emph{Stateful Dynamic Data Sharding Service}, \emph{Monitor}, \emph{Controller}, and \emph{Agent}. These components collaborate to dynamically allocate workloads, handle faults, and offer a comprehensive set of pre-defined straggler mitigation methods. Secondly, by encapsulating the intricate details of data allocation and fault tolerance within the framework, users could easily customize the straggler mitigation solutions and we propose two solutions to leverage the time periodicity and heterogeneity of stragglers to systematically alleviate different kinds of stragglers in the industrial scenarios. 

Our main contributions are:
\begin{enumerate}

\item AntDT firstly provides a self-adaptive framework to systematically resolve different kinds of stragglers in the industry-level distributed training but hides the messy details of data allocation and fault tolerance caused by different straggler mitigation methods.

\item  Based on the AntDT framework, we propose two straggler mitigation solutions as running examples to effectively alleviate the stragglers on both worker and server nodes, which outperforms other methods more than 3$\times$.

\item \textcolor{black}{Our comprehensive experiments and industrial deployment demonstrate the significant efficiency of AntDT in industry scenarios. It significantly decreases the training duration in the core recommendation scenario by 5$\times$.}
\end{enumerate}
\textcolor{black}{Furthermore, we plan to release the code for AntDT as an open-source project, pending approval from the company.}

\section{Problem Formulation}\label{sec:problem_formulation}
The section formalizes the straggler problem in the Parameter Server as an example and it could be easily extended to the AllReduce architecture. In one iteration of model training, given the global batch $B$, one worker $i$ of $n$ workers first calculates its local gradient based on the local batch $B_i$ and pushes it to the servers. Afterward, the servers aggregate the local gradients from all workers and update the parameters asynchronously or synchronously. We denote the entire duration of worker $i$ in one iteration as the batch processing time (BPT). It can be decomposed into three parts: \textbf{computation time on the worker node $T_i^w$} includes the forward and backpropagation time based on batch data $B_i$, which could be formalized as a function of batch data $F(B_i)$; \textcolor{black}{\textbf{computation time on the server node for worker $i$, namely $T_{i}^s$}, where the servers aggregate the local gradient from worker $i$ and update the model parameters; and the \textbf{communication time} between worker $i$ and servers, namely $T_i^m$, measures the communication time where worker pulls the latest parameters and push the local gradients for synchronization.} \textcolor{black}{Particularly, the computation time $T_i^s$ for specific worker $i$ is subject to the longest computation duration on $m$ servers, namely $\max_j\{ T_{ij}^s \}$.} Note that the global batch size $B$ in one iteration has to be fixed to ensure statistical performance.  

In the BSP mode of the Parameter Server, the whole training procedure can be formalized as the following optimization problem (it could be further \textcolor{black}{extended} to the AllReduce paradigm by ignoring the $T_{i}^s$ term). Given n workers and m servers, we want to minimize the longest batch processing time among all workers, i.e., 

\begin{small}
\begin{equation}
\begin{aligned}
\min \max_{i \in \{1, ..., n\}} \quad & T_{i}^{w} + T_{i}^{s} + T_{i}^{m}\\
\textrm{s.t.} \quad & \textcolor{black}{T_i^s = max_j\{ T_{ij}^s \}, \forall i \in \{1, ..., n\}, \forall j \in \{1, ..., m\}}  \\
  & T_i^w = F(B_i), \forall i \in \{1, ..., n\}    \\
  & \sum_{i=1}^n B_i = B  \\
\end{aligned}
\label{eq:eq1}
\end{equation}
\end{small}

\begin{table}[t]
\caption{\textcolor{black}{Summary of Important Notations.}}
\begin{tabular}{ll|lll}
\cline{1-4}
\textcolor{black}{$T_i^w$} & \textcolor{black}{worker comp. time of worker $i$} & \textcolor{black}{$v_{i}$} & \textcolor{black}{worker $i$'s speed} &  \\
\textcolor{black}{$T_{i}^s$} & \textcolor{black}{server comp. time of worker $i$} & \textcolor{black}{$L^{trans}$} & \textcolor{black}{short time window} &  \\
\textcolor{black}{$T_i^m$} & \textcolor{black}{comm. time of worker $i$} & \textcolor{black}{$L^{per}$} & \textcolor{black}{long time window} &  \\
\textcolor{black}{$B_i$} & \textcolor{black}{batch size of worker $i$} & \textcolor{black}{$\lambda$} & \textcolor{black}{slowness ratio} &  \\ \cline{1-4}
\label{tab:notation}
\end{tabular}
\end{table}

\section{Related Work}\label{sec:related_work}
\textcolor{black}{Stragglers have long been a thorny problem in traditional distributed computing systems, and many works have been proposed to mitigate them. Referring to classic works in the traditional distributed computing community~\cite{gill2020tails,ananthanarayanan2013effective,harlap2016addressing}, existing straggler mitigation methods in the ML community could be classified into:}

\textcolor{black}{\textbf{Load-Balancing} based approaches are commonly employed in task scheduling and data processing systems to ensure a fair distribution of workloads among all nodes~\cite{yang2018scheduling,acar2013scheduling,mitzenmacher2001power,dean2008mapreduce}. These approaches typically redistribute load from heavily loaded workers to lightly loaded ones during runtime. However, load-balancing based methods typically involve runtime load migration, which is inefficient for ML workloads. A ML job typically consists of thousands of short-lived iterations and these iteration tasks are non-idempotent and indivisible~\cite{chen2020semi,li2014scaling,harlap2016addressing}. The frequent coarse-grained load migration in traditional load-balancing-based methods may necessitate suspending the training process, which is inefficient. In the ML community, some works~\cite{zhou2020falcon,harlap2016addressing,zhou2019falcon} take similar methods. It requires suspending the training, and then proactively reassigning the workloads, especially for synchronous training (BSP). However, another method, referred to as ADJUSR\_BS~\cite{chen2020semi,yang2018adaptive,tyagi2020taming,ferdinand2020anytime}, could rebalance workloads by reducing the batch size of the slow worker and increasing the batch size of the fast worker without suspending the training.}  

\textcolor{black}{\textbf{Replication} based methods launch duplicate tasks of identified stragglers and only accept the results from the first finished tasks in a job~\cite{zaharia2008improving,dean2008mapreduce,da2003trading}.  Backup Workers (BACKUP\_WORKERS) ~\cite{chen2016revisiting, hanna2020adaptive,tandon2017gradient} utilized this method in the context of distributed machine learning by dropping a few slowest gradients (intermediate results for ML) in one iteration, which eliminates the straggling nodes on the worker side. While duplicating data can be seen as a method of sampling\cite{fernandez2018learning} in ML applications, discarding data from slower workers is unacceptable for many scenarios; otherwise, it harms the data integrity of the dataset and may compromise the statistical performance.  For instance, fraud detection heavily relies on ML techniques, where positive samples (fraudulent transactions) are significantly outnumbered by negative samples (normal transactions). Losing such positive samples is deemed unacceptable, particularly in financial applications~\cite{cao12titant,zhang2023framework,Zhou_Cao_Hu_Zhang_Chen_2023b}.
}

\textcolor{black}{\textbf{Scheduling} based methods involve scheduling the resources to produce the speculative copies of tasks~\cite{ananthanarayanan2013effective,ananthanarayanan2010reining,zaharia2008improving,ousterhout2013sparrow}. These methods are quite similar to replication-based methods but focus on allocating resources from cluster scheduler perspective. However, different from schedulers scheduling jobs at the sub-seconds level\cite{ananthanarayanan2010reining,ousterhout2013sparrow}, it is inefficient to clone short iterative tasks in ML jobs since the workers and servers need to synchronize across different iterations. Instead, the ML community focuses on a single job. These works ~\cite{or2020resource,wu2021elastic,peng2018optimus,harlap2017proteus}, referred to as KILL\_RESTART, temporarily suspend training, replace the slow nodes with newly launched ones, and then resume training from previously saved checkpoints. However, it incurs significant time overhead due to rescheduling and data integrity problems that data is duplicated or lost.}


\textcolor{black}{\textbf{Optimization} based methods are characterized to specific ML applications. It tries to solve the problem in model convergence from the ML optimization standpoint, but their impact on statistical performance is typically confined to a narrow range of training scenarios~\cite{dutta2018slow}. Especially, these works~\cite{qiao2021pollux,jiang2017heterogeneity,dutta2018slow}, referred to as ADJUST\_LR, dynamically adjust the learning rate of each worker in the ML optimizer by penalizing the lagging worker. This hopes to speed up the convergence of model training, which in turn improves the training efficiency.}

\begin{figure*}[t]
    \centering
\includegraphics[width=0.7\linewidth]{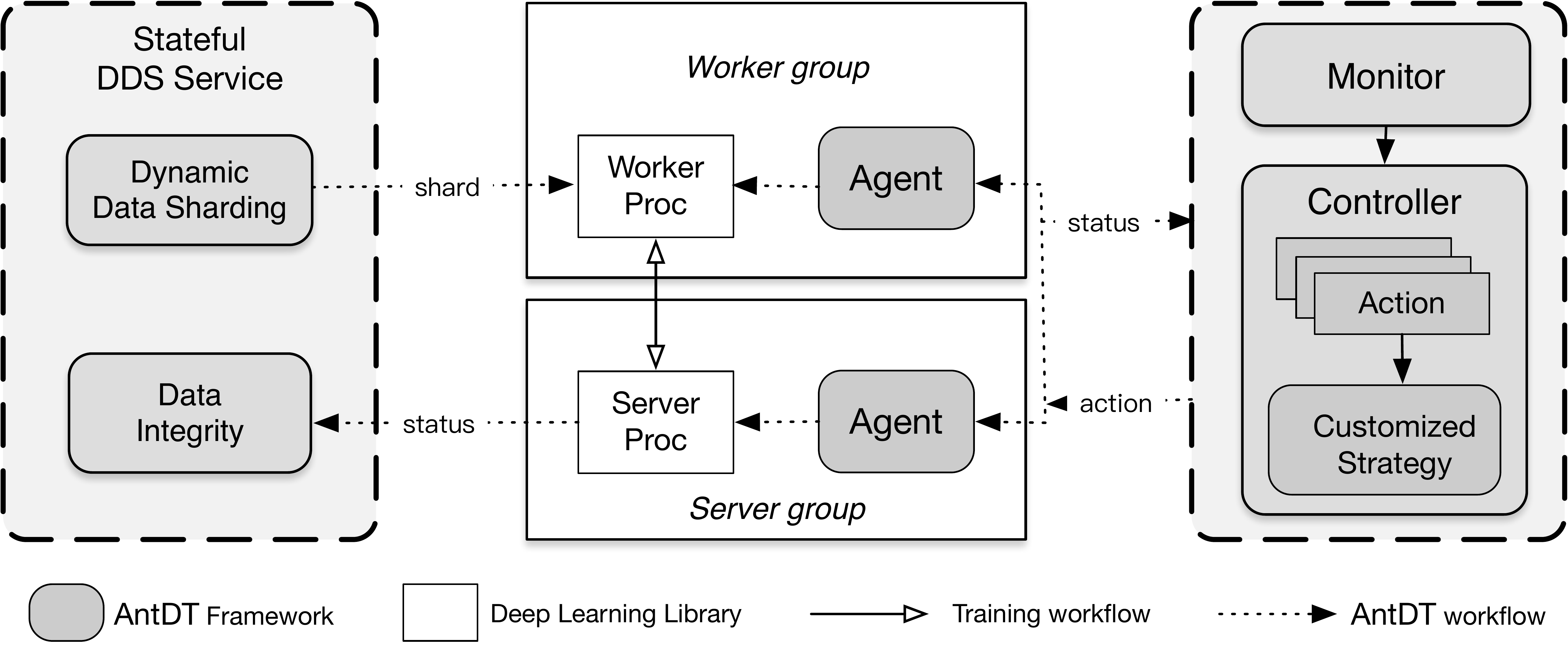} %
    \caption{Overview of AntDT Framework}%
    \label{fig:arch}%
\end{figure*}

\section{Chanllenges}
\label{sec:challenges}


\textcolor{black}{To completely address the straggler problem, we encounter three main challenges.}

\textcolor{black}{Firstly, it is necessary to minimize all three terms $T_{i}^w$, $T_{i}^s$, and $T_{i}^m$ at the same time formalized in the Eq. \ref{eq:eq1}, while the existing works only minimize part of these terms, failing to completely eliminate the stragglers.}  For example, load balancing-based methods could move the workloads from the slow workers to the fast workers to reduce the $T_i^w$. \textcolor{black}{However, it could not lessen the  $T_{i}^s$ and $T_{i}^m$  when there is resource contention since the worker node is responsible for most of the computation, and server nodes mainly account for IO-intensive storage and updating of the model parameters. Scheduling-based methods could reduce either $T_{i}^w$ or $T_{i}^s$ and $T_{i}^m$ terms. It kills the straggling nodes and relaunches a new node, which hopes to schedule to a leading node in terms of computing and network efficiency. Nevertheless, it is not practical to frequently utilize it when transient stragglers are epidemic due to the time cost~\cite{peng2018optimus, abadi2016tensorflow, torchelastic}. Therefore, we need to unify these works as straggler mitigation actions to effectively address all these stragglers in industrial training.}

Secondly, an adaptive data assignment strategy is crucial to ensure the scalability and compatibility of various straggler mitigation actions. Currently, these straggler mitigation methods could hardly be combined altogether to address the straggler problems. All these methods require different data allocation mechanisms to re-balance workloads among workers according to their specific straggler mitigation solution. \textcolor{black}{For example, adjusting batch size changes the batch size of each worker, replication-based methods require assigning more samples to leader workers, and traditional load-balancing methods~\cite{zhou2020falcon,harlap2016addressing,zhou2019falcon} proactively transfer the workloads among workers during runtime, which results in different data consumption among these workers.} These methods employ complicated data allocation strategies to actively transfer the data loads from slow workers to faster ones, which are incompatible with each other and inefficient in the industrial setting. Therefore, it requires an agile data allocation mechanism for the framework to utilize all these methods.

\textcolor{black}{Thirdly, Replication and scheduling-based methods may lose the samples during the training, compromising the data integrity composed by ``at-least-once" and ``at most once" semantics. Particularly, ``at-least-once" means that every sample should be used at least once, while ``at-most-once" means that every sample should not be used more than once for one epoch (one entire iteration of the whole dataset) during the training. Especially, ``at-least-once" semantic is more important than ``at-most-once" semantic in ML tasks for financial applications~\cite{cao12titant,zhang2023framework,Zhou_Cao_Hu_Zhang_Chen_2023,Zhou_Cao_Hu_Zhang_Chen_2023b}. Compromising the ``at-least-once" semantic, which loses samples or information for training is unacceptable as aforementioned in Section \ref{sec:related_work}. This requires considering the data integrity when applying these replication and scheduling-based straggler mitigation methods.}

Lastly, it is crucial to meticulously design an efficient fault tolerance mechanism, taking into account the agility of the data allocation. \textcolor{black}{The scheduling approach, namely KILL\_RESTART, includes killing the lagging nodes, and recovering the training states (mainly model states and IO states) from the periodically saved checkpoints during training into the newly launched node.} However, if KILL\_RESTART action is not properly implemented, it can compromise data integrity by either losing or duplicating the allocated data, thereby hindering statistical performance.  Additionally, the time cost of failover should be minimized to ensure that the KILL\_RESTART action does not impose more delay than the straggler itself.

\section{System}\label{sec:system}
\subsection{System Design Objectives}\label{sec:obj}
Based on our discussions, our objective is to design a distributed training framework that reduces the impact of stragglers while satisfying the following objectives:

\begin{itemize}
  \item \textcolor{black}{\textbf{Efficiency}. The framework should address various types of stragglers and speed up training in straggler-prone environments under different consistency models, e.g. BSP or ASP.}
  
  \item \textcolor{black}{\textbf{Data Integrity}. The framework should ensure the correctness of model training by maintaining ``at-least-once"  or ``at-most-once" semantics. Specifically, it is crucial to prioritize the ``at-least-once" semantics to prevent any loss of information.}

  \item \textcolor{black}{\textbf{Observability}. As discussed, stragglers own time periodicity. Monitoring statistics, such as throughput, is essential to detect such stragglers. However, different from online jobs, it has fewer requirements for latency, which only requires minute-level observability.}
  
  \item \textbf{Extensibility}. The framework should be able to employ a wide range of straggler mitigation techniques to address various types of stragglers. \textcolor{black}{Also, users could easily utilize these actions to customize the straggler mitigation solution to align with the specific circumstances of their cluster environment.}
  
  \item \textbf{Scalability}. The framework should be able to scale out to support large-scale distributed training in industry scenarios, such as hundreds of nodes.
\end{itemize}

\subsection{System Overview}
In this section, we outline the design of the entire system. The framework comprises four components: the \emph{Stateful Data Sharding Service (Stateful DDS}), \emph{AntDT Monitor}, \emph{Controller}, and \emph{Agent}, as illustrated in Fig. \ref{fig:arch}. It provides essential features like data allocation and fault tolerance with high efficiency and scalability while decoupling with specific straggler mitigation actions. Based on our framework, users could utilize pre-defined straggler mitigation solutions further shown in section \ref{sec:Straggler_Mitigation_Strategy} or easily customize the specific straggler mitigation solutions according to the practical circumstances. Firstly, the \emph{Stateful DDS} is designed to adaptively dispatch the data shards to worker nodes at a fine granularity. It also ensures data integrity and efficiently resumes the data shards when the worker terminates expectedly or unexpectedly. Secondly, \emph{AntDT Monitor} collects the information related to the straggler and listens to the pod notifications. Thirdly, the \emph{Controller} takes action according to the specific straggler mitigation solution. Lastly, the \emph{Agent} is deployed on each node to report the observed data to the \emph{Monitor} and execute the actual straggler mitigation action notified by the \emph{Controller}.

\subsection{Stateful Dynamic Data Sharding Service}
To provide an agile data allocation mechanism with data integrity under various straggler mitigation actions, we design and implement the \emph{Stateful Dynamic Data Sharding Service} (\emph{Stateful DDS}) in the AntDT framework. The main idea of Dynamic Data Sharding is to use a dynamic data partition strategy to assign corresponding data shards to workers with unbalanced computation capacity at a fine granularity.  Additionally, each shard owns a state that contributes to the data integrity in case of any node terminated. Then we could manage the training data at the shard level, including data assignment and data integrity.

\subsubsection{Dynamic Data Sharding}
\begin{figure}[t]%
    \centering
\includegraphics[width=1 \linewidth]{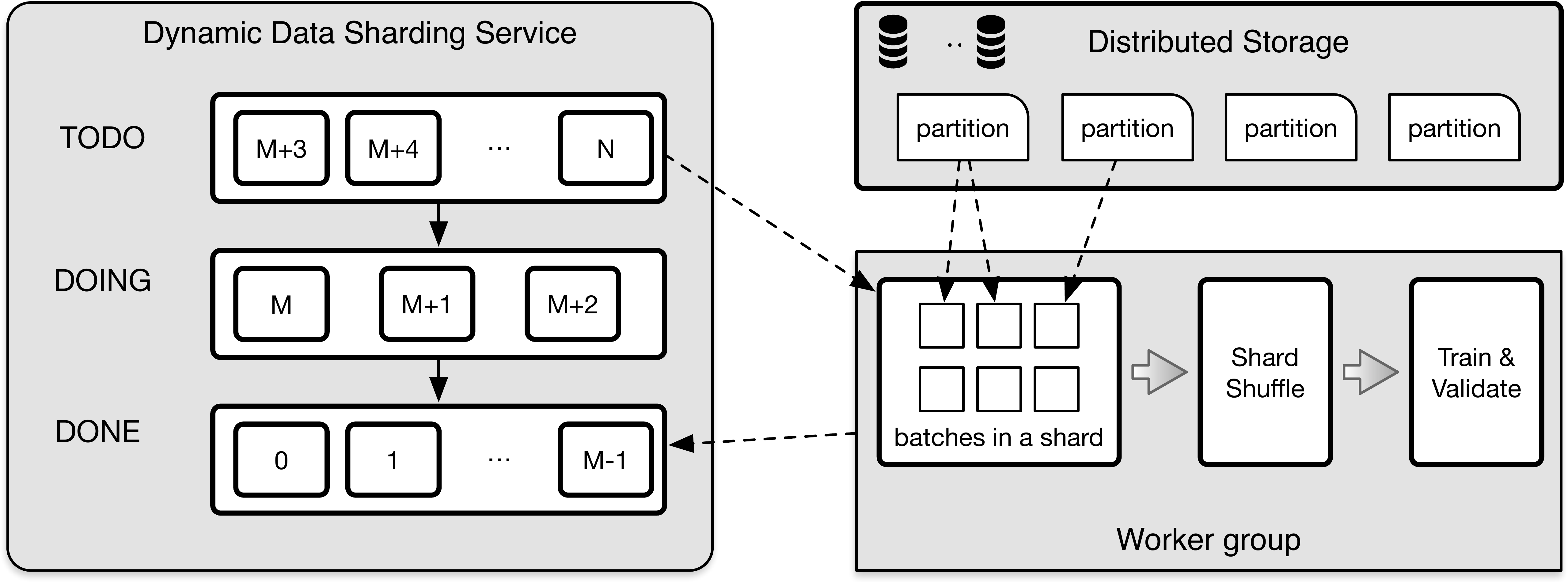} %
    \caption{Stateful Dynamic Data Sharding service (DDS)}%
    \label{fig:dds}%
\end{figure}

 The Dynamical Data Sharding service maintains a global message queue where workers fetch the shard for further training and report each shard's state, as shown in Fig. \ref{fig:dds}. In detail, the total $N$ training samples are split into $K$ data shards $D_i (i=1,..., K)$ where $K =\lceil \frac{N}{BM} \rceil $, $B$ is the global batch size and $M$ is the number of batches per shard. To save network cost, each shard only contains two integers, the start offset, and the length. A shard may contain several batches of record indexes. After that, all data shards are inserted into the queue for workers to consume. On the worker side, the worker fetches the data shard from the queue and reads samples from the data storage by mapping the offset and the length into actual input data such as file IO or SQL-like data. Eventually, the worker performs the model computation based on the given data. Additionally, a Shard Shuffler guarantees the order of the samples via shuffle between shards and shuffle among samples in the same shard.
Furthermore, the number of batches $M$ in each shard plays a vital role as the hyperparameter for granularity. For smaller $M$, we can have more precise control over workload distribution but at the cost of communication overhead between workers and the queue. Currently, we set the value to 100 by default and empirically increase it when the throughput of the model training is higher.

\subsubsection{Availability of Agile Data Allocation}
In contrast to the even data partition strategy, where each worker has the same amount of data to train, or complicated data allocation strategy as discussed in section \ref{sec:challenges}, the \emph{DDS} could agilely distribute more samples to fast workers while the stragglers naturally get less since the worker cannot fetch new shards until it has completed its current one. In this way, the \emph{DDS service} can allocate the data shards to the workers at a fine granularity according to the actual data consumption, rather than evenly \textcolor{black}{partitioning} the whole dataset. Compared with some works that proactively transfer the workloads in a complicated and communication-expensive way, the passive data allocation mechanism helps the training scale out to hundreds of nodes with low overhead in various load-balancing based methods.

\subsubsection{\textcolor{black}{Data Integrity}}
For data integrity, the \emph{DDS service} also hosts the state information of each shard to trace the completion status. These states are classified into three categories: ``TODO", ``DOING", and ``DONE". All of the state transitions are conducted by the \emph{DDS service}:

\begin{itemize}
  \item {\verb|"TODO" state|}: The initial state of all data shards. It means that the shard is ready for assignment;
  \item {\verb|"DOING" state|}: ``DOING" status represents that a worker is currently working on the given data shard and the shard won't be assigned to other workers;
  \item {\verb|"DONE" state|}: The shard is marked as ``DONE" after the worker successfully pushes local gradients to servers.
\end{itemize}
 
During the normal training, the worker fetches ``TODO" shards from the \emph{DDS service} in the very beginning, and the shard is marked as ``DOING", shown in Fig. \ref{fig:dds}. After that, the worker performs forward and backward propagation based on the pulled data shard. The worker reports the shard state after gradients have been pushed into servers, and \emph{DDS} marks these shards as ``DONE" when workers complete the shard. 

When any shard is lost in the case of BACKUP\_WORKERS or KILL\_RESTART, the assigned ``DOING" shard to the worker will be marked as ``TODO" by the \emph{DDS service}, and \emph{DDS} inserts the shard back into the end of the data queue. After the restart of the failed pod, the new worker pod requests the latest ``TODO" shard from \emph{DDS}, and the whole procedure guarantees ``at-least-once" semantics for the job execution. Furthermore, to guarantee ``at-most-once" semantics, it is necessary to set the number of batches in each shard as one, and all the data shards after the checkpoint need to be recomputed. This, unfortunately, results in a significant time cost. Nevertheless, we argue that relaxing the ``at-most-once" semantic by setting a relatively small number of $M$ is usually acceptable in distributed deep learning. \textcolor{black}{This is because slight data duplication can be viewed as a form of sampling.}

 \subsection{AntDT Monitor}
The \emph{AntDT Monitor} component is designed to collect observable information from different data sources for further straggler mitigation purposes in the \emph{Controller}. The \emph{AntDT Monitor} will periodically gather and aggregate three types of information for the straggler detection, including the application states, node states, and general information from other modules or third-party APIs in training. We provide APIs where the \emph{Controller} and other modules can access the aggregated data. The collected information is as follows:

\begin{itemize}
  \item {\verb|Node State|}: The node state contains the standard monitoring metrics such as node termination notifications and error codes via the Kubernetes controller. It is used for failover for \textit{KILL\_RESTART} action and fault tolerance purposes \footnote{Failover or fault tolerance is vital to KILL\_RESTART action and scalability in the industry. It enables the framework to scale out to hundreds of nodes where the probability of failures rockets up due to the breakdown of machines and networks (or job preemptions in the shared clusters)\cite{li2014scaling}. For practicality, we consider failovers from both KILL\_RESTART action and unexpected failures.}. We classify these ``errors" into retryable and unretryable errors. Typical retryable errors are proactive termination in KILL\_RESTART, network errors, and job eviction. Unretryable errors are usually configuration errors or programming errors from users, which should terminate the training job.  
  \item {\verb|Application State|}: The application state is the information related to the training speed of the deep learning process. This includes the batch processing time in the worker or server node and batch size information.
  \item {\verb|Third Party Information|}: Third party information is collected from other modules. For example, job pending time is collected from the cluster scheduler for the awareness of whether the cluster is busy or idle. 
\end{itemize}

\subsection{AntDT Controller}
The \emph{AntDT Controller} module is designed to hold different straggler mitigation solutions, which decouples with the framework. Users could easily customize their straggler mitigation solution based on the framework and the provided straggler mitigation actions set, neglecting the data allocation and fault tolerance problem shown in section \ref{sec:challenges}. We present two straggler mitigation solutions, AntDT-ND for non-dedicated clusters and AntDT-DD for dedicated clusters, as running examples. These solutions are further discussed in detail in section \ref{sec:Straggler_Mitigation_Strategy}. The \emph{AntDT Controller} module ingests the information from the \emph{Monitor} and sends the actions to the following \emph{AntDT Agent} according to the specific straggler mitigation solution. The difference between straggler mitigation actions and solutions is that these straggler mitigation actions have different time costs and gains, therefore the straggler mitigation solution takes the actions according to the characteristics of the cluster, such as the time periodicity of the stragglers.

\subsubsection{Straggler Mitigation Action Set}
We provide a pre-defined straggler mitigation action set in the \emph{AntDT Controller} for solutions to meet the need in different scenarios as displayed in Table \ref{table:Action_set}. \textcolor{black}{These actions are discussed in Section \ref{sec:related_work} and classified into two types: } 

\begin{itemize}
  \item {\verb|Node Action|}: One type of action is the Node Action like \textit{KILL\_RESTART}. It is independent of actions on other nodes which does not require synchronization.
  \item {\verb|Global Action|}:  Another type of action is the Global Action like \textit{ADJUST\_BS}, \textit{BACKUP\_WORKERS} and \textit{ADJUST\_LR}, which requires synchronization among nodes. 
\end{itemize}

\begin{table}[t]
\caption{Straggler Mitigation Action}
\centering
\begin{tabular}{ c|c|c } 
\hline
\textbf{\textcolor{black}{Action} }& \textcolor{black}{\textbf{Type} }& \textcolor{black}{ \textbf{Description} }\\
 \hline
 \textit{\textcolor{black}{ADJUST\_BS} }& \textcolor{black}{Load-balancing }& \textcolor{black}{Adjust the batch size  }\\
 \textit{\textcolor{black}{BACKUP\_WORKERS} }& \textcolor{black}{Replication }& \textcolor{black}{Use backup workers  
 }\\
 \textit{\textcolor{black}{KILL\_RESTART} }& \textcolor{black}{Scheduling }& \textcolor{black}{Kill and restart the node }\\
 \textit{\textcolor{black}{ADJUST\_LR} }& \textcolor{black}{Optimization} & \textcolor{black}{Adjust the learning rate }\\
 \textit{\textcolor{black}{NONE} }& \textcolor{black}{/ }& \textcolor{black}{Dummy action }\\
 \hline
\end{tabular}
\label{table:Action_set}
\end{table}

\subsubsection{Time-cost and Gains of Actions}
These straggler mitigation actions have various time costs and gains. On one hand, the \textit{KILL\_RESTART} action is crucial in resolving severe \textit{persistent stragglers} caused by resource contention on worker nodes, where stragglers can be up to eight times slower than other nodes. This action also effectively addresses computing and communication stragglers on server nodes by reducing $T_{i}^s$ and  $T_{i}^m$. However, this action is very time-consuming. One part of the time cost comes from the scheduling, which includes new node initialization and pending time in the cluster scheduling queue. The pending time is negligible when the cluster is idle but could be dozens of minutes at the peak period. Another part of the time cost derives from the deep learning application. It includes rebuilding the communication world, restoring from the last checkpoint, reconstructing the computation graph, and recomputing the samples after the previous checkpoint to ensure statistical performance. This phase usually takes from several minutes to dozens of minutes. On the other hand, load-balancing or replication-based methods are effective in alleviating stragglers that come from transient resource contention and hardware heterogeneity with low time costs. However, these methods are ineffective in minimizing the $T_{i}^s$ and $T_{i}^m$, compared with the \textit{KILL\_RESTART} action.

\subsubsection{Reducing the Time-cost of Worker \textit{KILL\_RESTART} }
Additionally, we optimize the \textit{KILL\_RESTART} phase based on the \emph{Stateful DDS} service on the worker side. In contrast to server failovers, the latest model parameters are still held on the servers after the failover on the worker side. In such cases, with the reading states from \emph{Stateful DDS}, we only need to recompute the data shards after the failover, as opposed to the data integrity approach in mainstream libraries, where the system restores from checkpoints and recomputes all intermediate results between checkpoints. This approach results in a significant reduction in the time-cost of the \textit{KILL\_RESTART} action on the worker side.

 \begin{figure}[t]%
    \centering
\includegraphics[width=1 \linewidth]{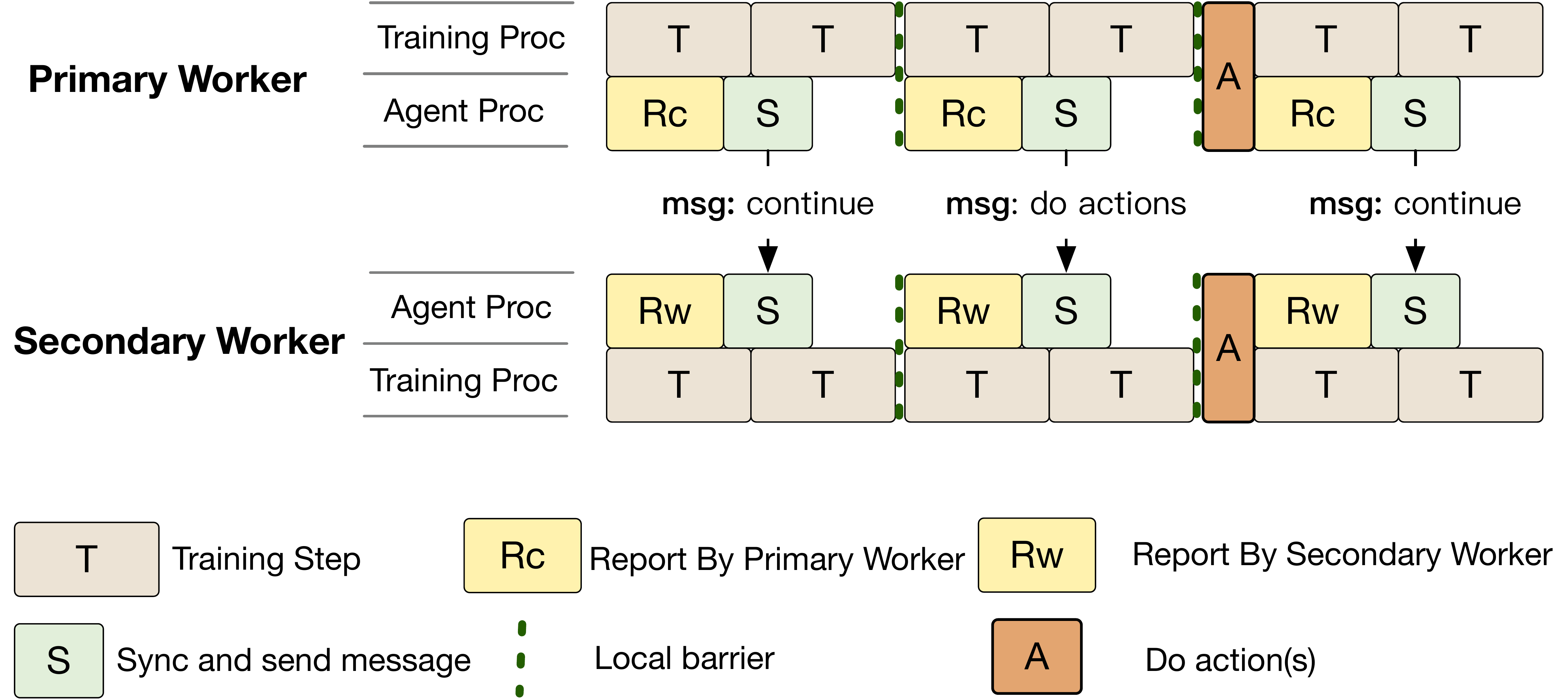} %
    \caption{Synchronization Mechanism}%
    \label{fig:sync-mechanism}%
\end{figure}

\subsection{AntDT Agent}\label{sec:sys_agent}
 The \emph{AntDT Agent} (process) is designed to collect information for the \emph{Monitor} and execute the actions from the \emph{Controller} asynchronously. It is deployed in each worker or server node. On one hand, it asynchronously collects and pushes the corresponding application and node information to the \emph{Monitor}. On another hand, the \emph{Agent} executes the straggler mitigation action notified by the \emph{Controller}. The \emph{Agent} process communicates with the worker or server process via the environment variables in our implementations. 

Additionally, we design a synchronization mechanism for global actions like \textit{ADJUST\_BS}, which requires synchronization primitives to make each worker execute the command in the same iteration. It also lowers the time cost of synchronization during the phase. As depicted in Fig. \ref{fig:sync-mechanism}, the \emph{Agent} first reports the state information to the \emph{Monitor} periodically (every two iterations in Fig. \ref{fig:sync-mechanism}). \textcolor{black}{The \emph{Primary Agent}, \textcolor{black}{which is randomly elected similar to the primary worker}, receives a response message from the \emph{Controller}. Then the \textcolor{black}{\emph{Primary Agent}} broadcasts the message to all \textcolor{black}{\emph{Secondary Agents}}. The training process and the \emph{Agent} process get synchronized by a local barrier. After that, except for the \textit{None} action, all the workers execute the same action in the next iteration. In this procedure, the overhead of the local barrier is almost negligible because the report duration and synchronization between the training processes and \emph{Agents} are minor. These communications typically consist of bytes-level signals.}

\section{Straggler Mitigation \textcolor{black}{Solution}}\label{sec:Straggler_Mitigation_Strategy}
In this section, based on the AntDT framework, we design and implement two straggler mitigation solutions as running examples to show how to collaborate with the framework to solve the straggler's problems according to the cluster status at Ant Group. These solutions could be easily extended to other clusters in practice. The first solution, AntDT-ND (for the Non-dedicated cluster), leverages the time periodicity of stragglers in the non-dedicated cluster and takes specific actions to alleviate the stragglers and improve training efficiency. It also solves the straggler problems on both workers and servers. \textcolor{black}{Secondly, we present  AntDT-DD (for the dedicated cluster), which further optimizes the \textit{ADJUST\_BS} action to maximize the training speed in the dedicated GPU clusters.}

\subsection{Straggler Mitigation Solution in Non-dedicated Cluster} \label{sec:strategy_antnd}
As explained in section \ref{sec:introduction}, distributed training in non-dedicated clusters may suffer from both transient stragglers and persistent stragglers with different time characteristics. We take the parameter server architecture as an example since stragglers occur on both the worker and server sides. In the AntDT-ND, we not only leverage the time periodicity of stragglers in the non-dedicated cluster to reduce the time cost of straggler mitigation action, but also solve both stragglers on workers and servers to further improve training efficiency. Specifically, on the worker side, we take the lightweight \textit{ADJUST\_BS} action to alleviate the transient stragglers and cautiously take the heavyweight \textit{KILL\_RESTART} to remove the persistent straggler, which minimizes the $T_i^w$. On the server side, we take \textit{KILL\_RESTART} action to minimize $T_i^s$ and $T_i^m$. \textcolor{black}{It is worth noting that the considerations of data allocation, data integrity, and fault tolerance are already addressed by the AntDT framework. Therefore, in the following procedures, we will focus on explaining how AntDT-ND specifically tackles the straggler problem on both workers and servers.}

\textbf{AntDT-ND Solution for Worker Stragglers:}

\begin{enumerate}
\item{\textcolor{black}{\textbf{Initialization}: We  evenly initialize the batch size $\{B_i\}_{i=1}^n$ assigned for all $n$ workers in the first iteration, where $B_i = B / n$. $B$ is the global batch size fixed in each iteration.}}

\item{\textcolor{black}{\textbf{Collecting Statistics via \emph{Monitor}}: We collect and average the batch processing time (BPT) $T_i^{w}$ for each worker $i$ from the \emph{AntDT Monitor} over two short-term and long-term sliding time windows, $L^{trans}$ or $L^{per}$ (in minutes) in recent iterations. The resulting BPTs are $\bar{T}_{i}^{trans}$ and  $\bar{T}_{i}^{per}$. $L^{trans}$  and $L^{per}$  are hyper-parameters, e.g. 5 minutes or 30 minutes, which help capture either transient or persistent straggler patterns. }}

\begin{figure}
\begin{minipage}[t]{0.45\linewidth}
\includegraphics[width=\linewidth]{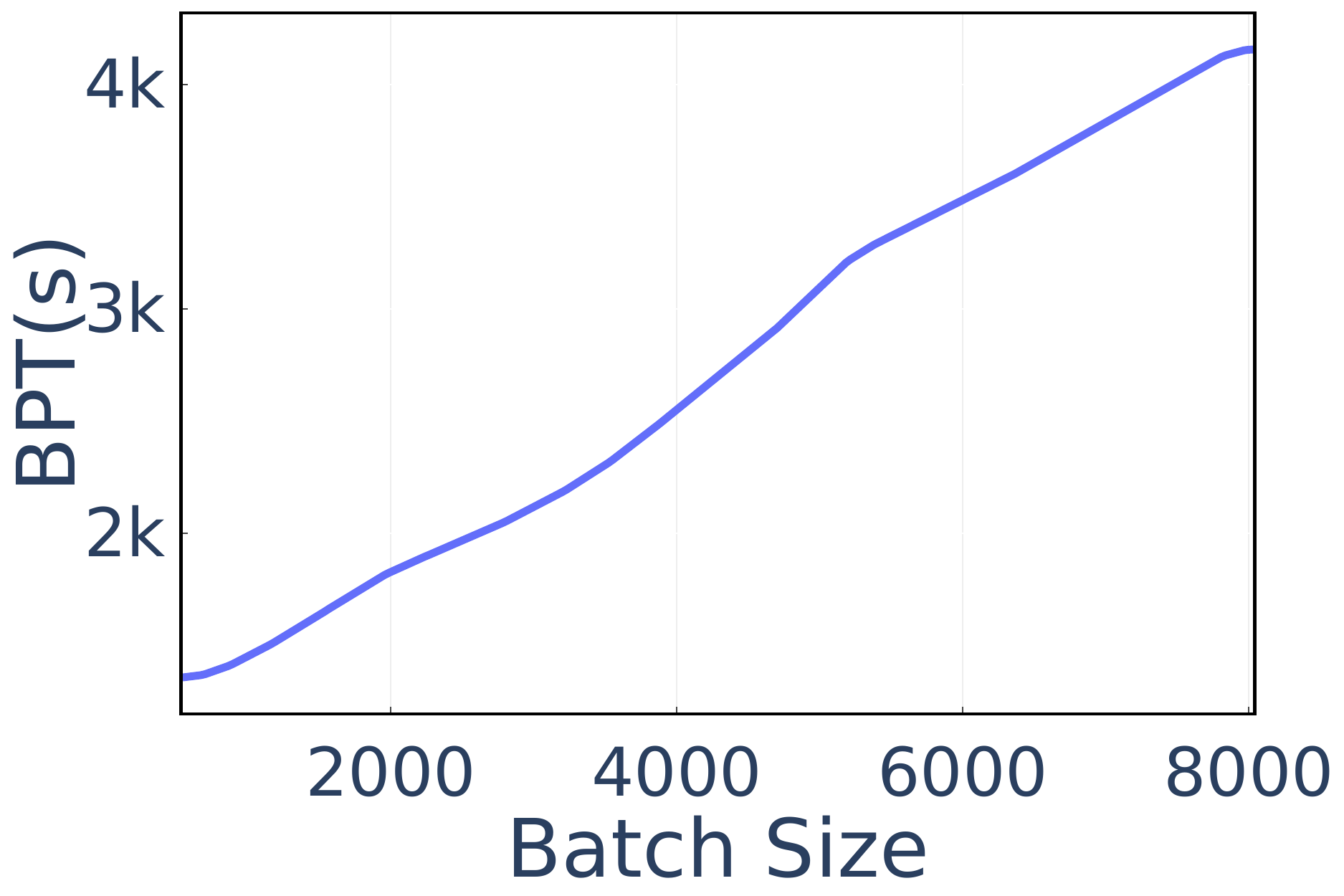}
\caption{BPT varying batch size in CPU cluster}
\label{fig:bs-cpu}%
\end{minipage}
\hfill
\begin{minipage}[t]{0.45\linewidth}
\includegraphics[width=\linewidth]{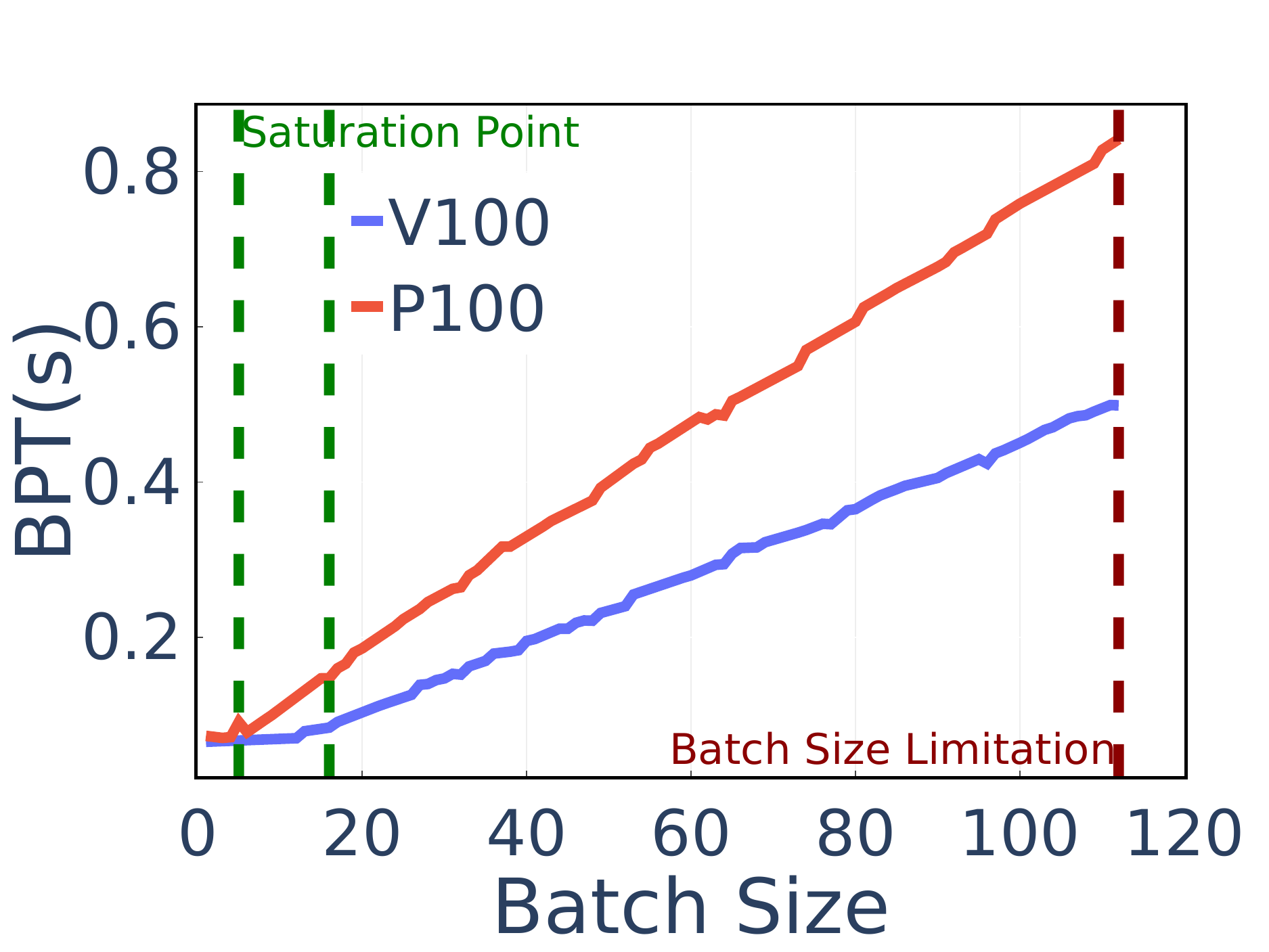}
\caption{BPT varying batch size in GPU cluster}
\label{fig:bs-gpu}%
\end{minipage}%
\end{figure}

\item{\textcolor{black}{\textbf{Solving Transient Stragglers via \emph{Controller}}: The worker $i$ is a \textit{Transient Straggler} if its short-term BPT, $\bar{T}_{i}^{trans}$, is much larger than average BPT $\bar{T}^{trans}$ over all workers ($\bar{T}_{i}^{trans} \geq \lambda \cdot \bar{T}^{trans}$). $\lambda$ is a heuristic factor\footnote{\textcolor{black}{In our practice, $\lambda$ is typically set to a value larger than 1.3 and is influenced by the cluster status, including the idleness of workloads. For instance, a value of 1.3 implies that the node is approximately 30\% slower than other nodes.}} that indicates the relative slowness ratio of a node compared to the average BPT over all nodes. We take \textit{ADJUST\_BS} action if the transient straggler is detected.}} 

\textcolor{black}{To attain the batch size of each worker to adjust, we assume the computation duration of the CPU device is linear to the batch size, which is validated in Fig. \ref{fig:bs-cpu}. We aggregate the throughput (batch size divided by the BPT) of workers over the short-term sliding window $L^{trans}$ to get $v_{i}$ (samples/sec). Mathematically, $v_{i} = \frac{1}{\abs{S_{L^{trans}}}} \cdot \sum_{S_{L^{trans}}}^{} \frac{B_i}{\bar{T}_{i}^{trans}}$.  To further estimate the BPT $T_i^w$ on the worker side, we could use the Equation $F(B_i) = B_i / v_{i}$. Ignoring $T_i^s$ and $T_i^m$, Eq. \ref{eq:eq1} could be simplified as:}

\begin{equation}
\begin{aligned}
\min \max_{i \in \{1, ..., n\}} \quad & T_{i}^{w}\\
\textrm{s.t.} \quad & T_i^w = F(B_i), \forall i \in \{1, ..., n\}    \\
  & \sum_{i=1}^n B_i = B  \\
\label{formula2}
\end{aligned}
\end{equation}

Using a latent variable $z$, the non-linear MinMax problem could be simplified into a Mixed-integer programming problem: 
\begin{equation}
\begin{aligned}
\min \quad &z \\
\textrm{s.t.} \quad & z \geq B_i /  v_{i}, \forall i \in \{1, ..., n\}  \\
  & \sum_{i=1}^n B_i = B  \\
\label{formula3}
\end{aligned}
\end{equation}

The Eq. \ref{formula3} could be solved easily, given global batch size $B$ and workers' throughput  $v_i$. The resultant batch size for each worker minimizes $T_i^w$ and the \emph{Controller} returns the ADJUST\_BS action to the \emph{Agent} with a series of batch size $\{B_i\}_{i=1}^n$ for the next iteration.

\item{\textcolor{black}{\textbf{Solving Persistent Stragglers via \emph{Controller}}: The worker $i$ is a  \textit{Persistent Straggler} if its long-term BPT, $\bar{T}_i^{per}$, significantly exceeds the average BPT $\bar{T}^{per}$ over all workers ($\bar{T}_{i}^{per} \geq \lambda \cdot \bar{T}^{per}$).  We take the \textit{KILL\_RESTART} action, if the persistent straggler is detected and the cluster is not busy (obtained from the \emph{Monitor}) when the job pending time is acceptable. The \emph{Controller} will then send the command to \emph{Agent} and reduce the $T_{i}^w$.}}

\item{\textcolor{black}{\textbf{No Straggler Case:} Lastly, the \emph{Controller} sends the dummy action \textit{None} to the \emph{Agent} if no transient or persistent straggler is detected. The whole procedure from 2) to 5) repeats until the end of training. }}\newline
\end{enumerate}

\textbf{AntDT-ND Solution for Server Stragglers:}

\begin{enumerate}

\item{\textcolor{black}{\textbf{Collecting Statistics via \emph{Monitor}}: We collect and average the BPT $T_i^{s}$ from each server $i$ from the AntDT \emph{Monitor} over the sliding windows $L^{per}$ in recent iterations, to get $\bar{T}_{i}^{per}$.}}

\item{\textcolor{black}{\textbf{Solving Persistent Stragglers via \emph{Controller}}: The server $i$ is a Persistent Straggler if its longer-term BPT $\bar{T}_{i}^{per}$ is much larger than average BPT $\bar{T}^{per}$ over all servers ($\bar{T}_{i}^{per} \geq \lambda \cdot \bar{T}^{per}$). The \emph{Controller} returns the \textit{KILL\_RESTART} action and reduces the $T_{i}^s$ as well as $T_{i}^m$ shown in Eq. \ref{eq:eq1}.}}

\item{\textcolor{black}{\textbf{No Straggler Case:} Similarly, the \emph{Controller} sends the dummy action \textit{None} to the \emph{Agent} if no persistent straggler is detected. The whole procedure from 1) to 3) repeats until the end of training.}} \newline

\end{enumerate}

\subsection{Straggler Mitigation Solution in Dedicated Cluster}\label{sec:strategy_antdd}
\begin{figure}[t]%
    \centering
\includegraphics[width=0.8 \linewidth]{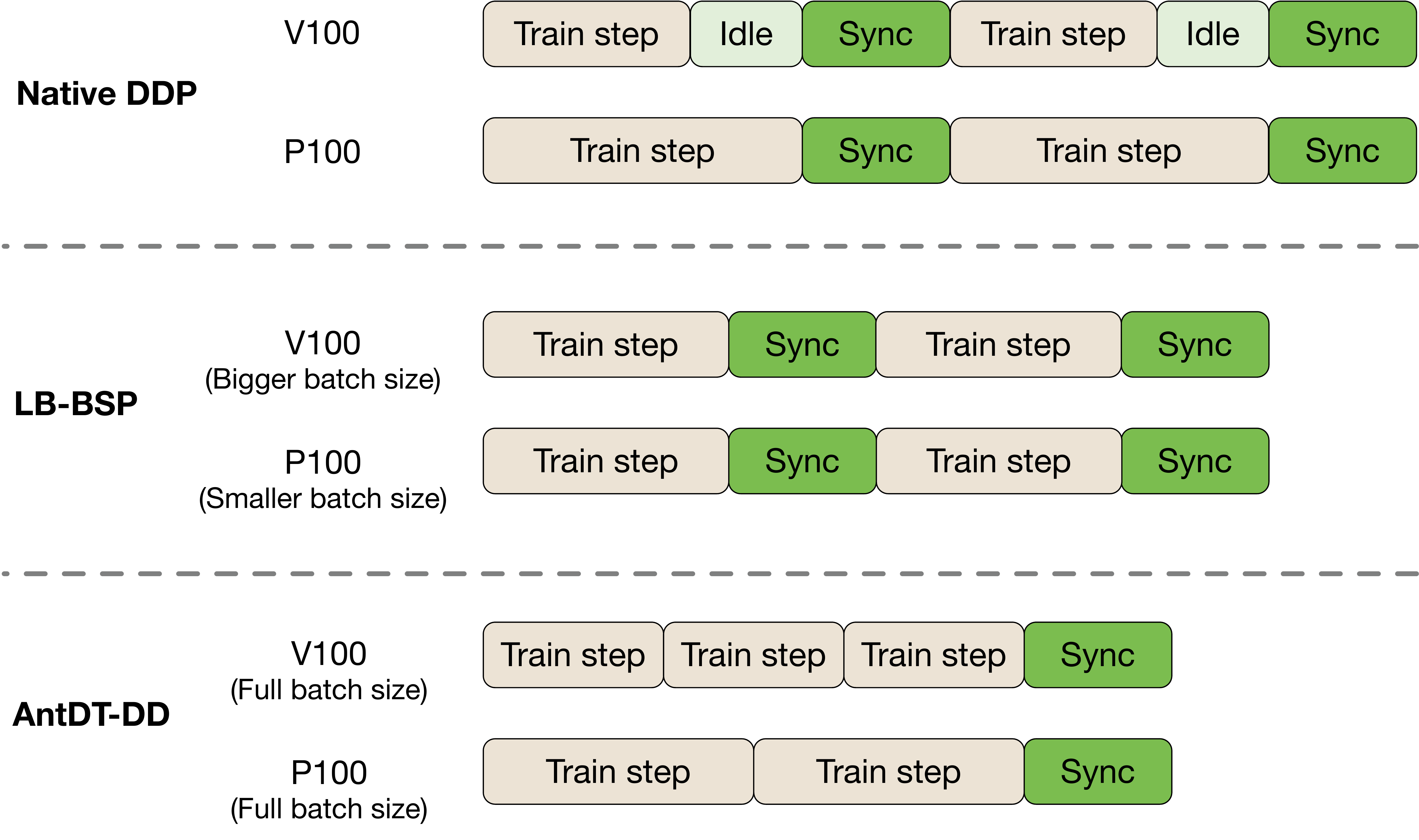}
    \caption{Gantt charts to compare the procedures of DDP (Distributed Data Parallel), LB-BSP, and AntDT-DD in the AllReduce.}
    \label{fig:opt-ddp}%
\end{figure}

\textcolor{black}{This section illustrates another solution, AntDT-DD, as a running example in the dedicated cluster. It further improves the ADJUST\_BS action in such clusters, where stragglers occur due to hardware heterogeneity, known as deterministic stragglers (as explained in Section \ref{sec:introduction}).} The batch size adjustment method like LB-BSP~\cite{chen2020semi} can effectively reduce the discrepancy in computation time across various GPU devices. However, there are still some drawbacks. As shown in Fig. \ref{fig:opt-ddp}, while LB-BSP reduces the idle time of advanced devices by minimizing the BPT gap between fast and slow devices through batch size adjustment, it still wastes the computing capacity of slower devices as their batch size is reduced and GPU memory is left unused.

To address these drawbacks, we propose the usage of Gradient Accumulation ~\cite{lin2017deep} and the Mixed-integer optimization method to further maximize the throughput of each GPU device. With this approach, the original batch size is split into several mini-batches, which are then computed sequentially before synchronizing the model weights. Our approach enables all devices to utilize their full batch size, maximizing throughput by computing several sequential batches to minimize the time gap before the next synchronization.

\textcolor{black}{After introducing the gradient accumulation, we further the Eq. \ref{formula2} as follows}. Assuming that we get $k$ different series of GPU devices, each device owns different numbers: $n_1, ..., n_k$. Let the number of gradient accumulations be $C_1, ..., C_k$. The $\hat{C}^{min}$ is the minimum number of gradient accumulation given by users (usually 1), and $\hat{C}^{max}$ is the maximum (e.g., 5). The $\hat{B}_i^{min}$ is the saturation point\footnote{Saturation point is where the batch computation time is constant even if the batch is pretty small shown in Fig. \ref{fig:bs-gpu}.} and $\hat{B}_i^{max}$ is the batch size limitation\footnote{The batch size limitation is the batch size where GPU memory usage is 95\%; otherwise it will run out of GPU memory (OOM).}. Then the problem could be formalized into the following integer MinMax optimization to minimize the batch processing duration before synchronization.

\begin{equation}
\begin{aligned}
\min \max_{i \in \{1, ..., k\}} \quad & C_i \cdot \frac{B_i}{v_{i}} \\
\textrm{s.t.} \quad & \sum_{i=1}^n n_i C_i  B_i = B  \\
  & \hat{B}_i^{min}  \leq B_i \leq \hat{B}_i^{max}, \forall i \in \{1, ..., k\} \\
  & \hat{C}^{min} \leq C_i \leq \hat{C}^{max}, \forall i \in \{1, ..., k\} \\
\end{aligned}
\label{formula4}
\end{equation}

\textbf{AntDT-DD Policy for Workers in Dedicated Cluster:}
\begin{enumerate}

\item \textcolor{black}{\textbf{Collecting statistics via \emph{Monitor}}: In the beginning, we collect the BPT and compute the throughput of each worker $v_{i}$. Also, $\hat{B_i}^{min}$ and $\hat{B_i}^{max}$ could be easily obtained by varying the batch size.}

\item{\textcolor{black}{\textbf{Solving Deterministic Stragglers via \emph{Controller}}: After that, we solve Eq. \ref{formula4} using the latent variable shown in Eq. \ref{formula3} to obtain the number of batch size $\{B_i\}_{i=1}^n$ and gradient accumulation $\{C_i\}_{i=1}^n$ for the next iteration. Subsequently, the \emph{Controller} returns the \textit{ADJUST\_BS} action to the \emph{Agent}. Note that adjusting the batch size only needs to be performed once since these stragglers are deterministic in the dedicated cluster.}}
\end{enumerate}

\section{Experiment and discussion}\label{sec:exp}
In this section, we introduce several experiments to evaluate the AntDT framework and solutions, focusing on the following questions:

\begin{enumerate}
\item How effective is the AntDT-ND in the non-dedicated compared with SOTA methods? (Q1)

\item How effective is the AntDT-DD in the dedicated cluster compared with SOTA methods? (Q2)

\item How does the AntDT framework perform concerning the agility of data allocation, data integrity, and reducing the time-cost of KILL\_RESTART action? (Q3)

\item Can the AntDT scale out well with growing computing nodes in the industrial-level distributed training? (Q4)
\end{enumerate}

\subsection{Experiment Setup}\label{sec:exp_setup}

\subsubsection{Cluster Setup}
We employ three types of clusters from the Ant Group Cloud for the evaluation experiments:

\begin{itemize}
  \item {\verb|Cluster-A|} is a dedicated CPU cluster that contains 20 workers and 8 servers. Each worker owns 16 CPU cores and 32 GB RAM; each server occupies 4 CPU cores and 24 GB RAM.
  \item {\verb|Cluster-B|} is a dedicated GPU cluster with 8 nodes which consists of four Tesla V100 GPUs and four Tesla P100 GPUs without NVLinks and each node is connected via 100Gbps bandwidth.
  \item {\verb|Cluster-C|} is a non-dedicated CPU cluster with workers having 16 CPU cores and 32 GB RAM, and servers having 12 CPU cores and 16 GB RAM. It has three node scale settings: small, medium, and large, consisting of 30, 60, and 90 worker nodes respectively, along with 12, 24, and 36 corresponding server nodes.

\end{itemize}

\subsubsection{Workload}
We evaluate AntDT by three typical workloads over two open-source benchmarks and one Ant Group production dataset in the TensorFlow Parameter Server and PyTorch DDP/AllReduce strategy. Firstly, we train the XDeepFM~\cite{lian2018xdeepfm} on three epochs of public Criteo dataset\cite{criteoailab} (containing 45 million user click records) in Cluster-A to systematically assess the performance of AntDT-ND. The global batch size $B$ sets to 81920, and the local batch size of each worker is 4096 on average. Secondly, we train the ResNet-101~\cite{he2016deep} and Moblienets~\cite{howard2017mobilenets} with one epoch of ImageNet~\cite{deng2009imagenet}(including 1.28 million images) in Cluster-B to evaluate the AntDT-DD. The global batch size  $B$ is set to 768. Lastly, we evaluate the framework using an in-house deep learning model consisting of several transformer blocks with one epoch of the Ant Group dataset (containing 2.7 billion samples) in cluster-C to verify the scalability of AntDT-ND at the industry level. The global batch size $B$ is set to 30720.

\subsubsection{Comparison}
\textcolor{black}{We conduct a comparison between two AntDT solutions and multiple baselines\footnote{Except for the ASP using the even data partition strategy for comparisons, all other methods utilize the \emph{Stateful DDS} as the data allocation strategy.} in either BSP or ASP mode. We include native training and existing single straggler mitigation methods like load-balancing and replication-based approaches for comparative analysis. However, we exclude the optimization-based method ADJUST\_LR, since it is closely linked to model accuracy, which is unfair to other methods.}
\begin{itemize}
\item {\verb|BSP|}~\cite{abadi2016tensorflow}: \textcolor{black}{It is a baseline that uses the native BSP training in the TensorFlow Parameter Server strategy.}

\item {\verb|Backup Workers(BW)|}: \textcolor{black}{We utilize the replication-based method, Sync-OPT~\cite{chen2016revisiting}, as the baseline. It disregards gradients from $b$ slowest workers in each iteration, but we leverage the \emph{Stateful DDS} to put back the abandoned samples, which ensures the data integrity. }

\item {\verb|LB-BSP|}~\cite{chen2020semi}: \textcolor{black}{For the load-balancing method, we use the LB-BSP's batch size updating algorithm in both the CPU cluster and the GPU cluster.}

\item {\verb|ASP|}~\cite{abadi2016tensorflow}: 
\textcolor{black}{It is a baseline that uses the native ASP mode in TensorFlow under the even data partition strategy. The job completion time is decided by the slowest worker.}

\item {\verb|ASP-DDS|}: ASP-DDS employs the Dynamic Data Sharding service as the data allocation strategy for fairness.

\item {\verb|AntDT-ND|}: In ASP training, AntDT-ND only takes the \textit{KILL\_RESTART} action. In BSP training, the AntDT-ND implements the solution following Section \ref{sec:strategy_antnd}. 

\item {\verb|DDP|}~\cite{paszke2019pytorch}: The native AllReduce training in PyTorch.

\item {\verb|AntDT-DD|}: AntDT-DD implements the solution following Section \ref{sec:strategy_antdd} in PyTorch. 
\end{itemize}

\subsubsection{Straggler Pattern} 
We produce the synthetic straggler patterns in the experiment following FlexRR~\cite{harlap2016addressing} since the cause of naturally occurring stragglers in the non-dedicated cluster is hard to control. We insert the following straggler patterns in the dedicated CPU cluster to evaluate how the framework performs under the straggler effects.  Also, the \textit{Straggler Intensity}, ranging from 0 to 1,  denotes the strength of how a node is disturbed. We emulate the time-varying resource contention by inserting the \textit{sleep} command into the training procedure in the worker or server threads. Generally, the delay duration could be formulated as $T_{delay} = SleepDuration \times Intensity$ with a certain probability. The $SleepDuration$ is the timespan (seconds) that is delayed in one iteration.

{\verb|Transient Stragglers|}: We insert $T_{delay}$ lasting over 15 minutes every 30 minutes into the workers with the probability of 0.3 in the training phase. That is to say that three of the ten nodes are likely to be disturbed. The sleep commands will add an extra $t$ seconds for each node to the usual batch processing time. We set the $SleepDuration$ to 1.5 seconds and raise the $Intensity$ from 0.1 to 0.8. 

{\verb|Persistent Stragglers|}: We emulate the long-term resource contention by inserting the constant straggler into servers or workers from the start to the end of the training. We constantly set the $T_{delay}$ to 4 seconds.

\subsubsection{Hyperparameters}\label{sec:hyperparameters}
For the granularity of the data shards, we set $M$ to 100, representing 100 batches in each data shard. For the frequency of data collection and execution, the \emph{Agent} reports states every 10 iterations, and the \emph{Monitor} aggregates and takes actions every 5 minutes. \textcolor{black}{For the AntDT-ND solution, we have set the relative slowness ratio $\lambda$ to 1.5, along with the sliding window sizes $L^{trans}$ and $L^{per}$ set to 5 and 10 minutes.}
 
\subsubsection{Evaluation Metrics}
Several evaluation metrics are used to evaluate the performance of the framework. Firstly, the job completion time (JCT), which measures the training time from the start to the end of the training, is adopted to evaluate the training efficiency. Secondly, we use the batch processing time (BPT) on both worker and server sides to measure the efficiency of nodes in one iteration. Lastly, we assess the statistical performance using AUC (Area under ROC Curve).

\subsection{Evaluation of AntDT-ND (for Q1)} \label{sec:exp_q1}
This subsection evaluates whether AntDT's solution in the non-dedicated cluster (AntDT-ND) could address both worker and server stragglers and improve the training speed compared with SOTA methods.

\begin{figure}
\begin{minipage}[t]{0.45\linewidth}
\includegraphics[width=\linewidth]{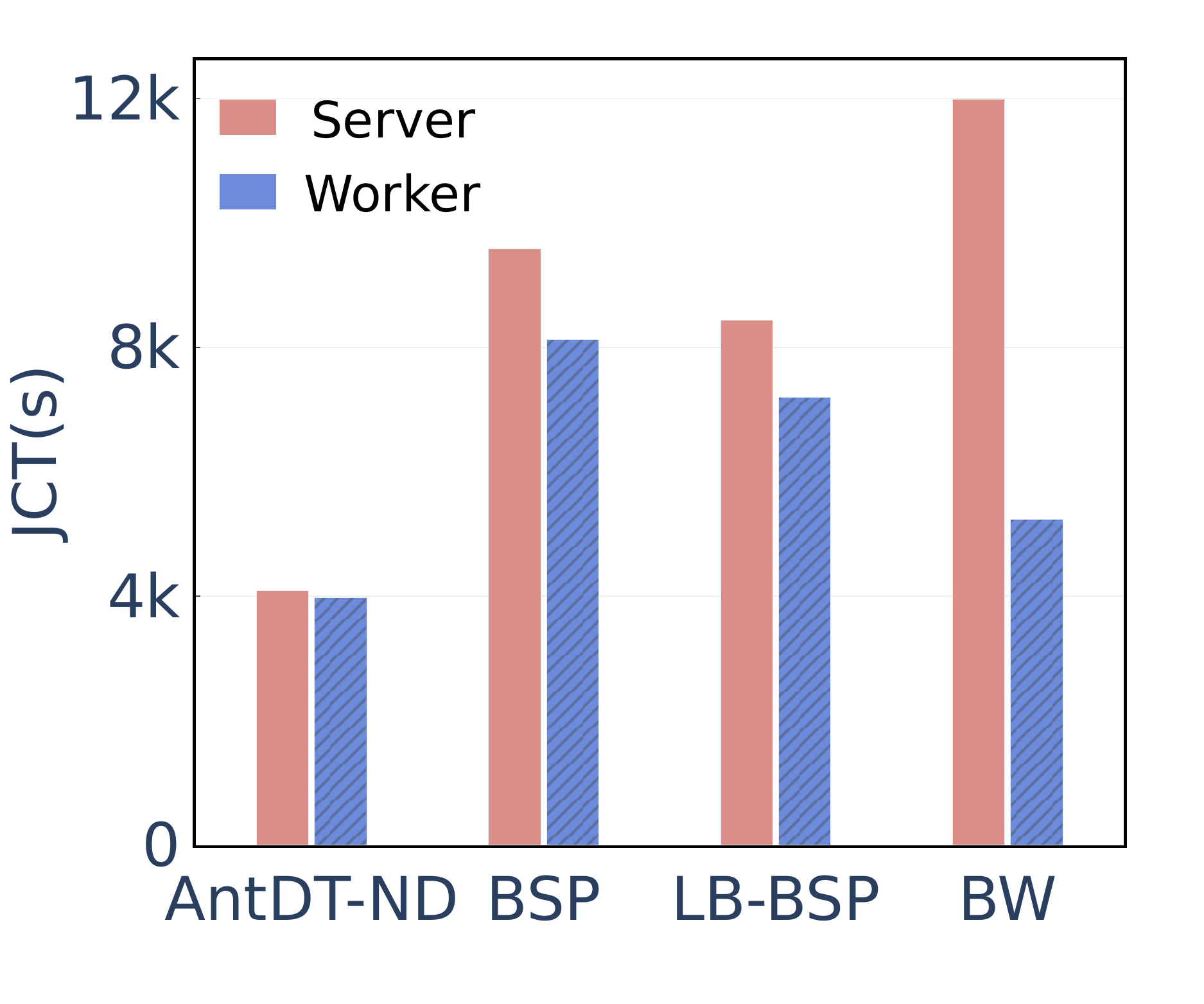}
\caption{JCT for methods in BSP training. The black bar is JCT in worker stragglers and the red bar is JCT in server stragglers.}
\label{fig:jct-syn-comparision}
\end{minipage}
\hfill
\begin{minipage}[t]{0.45\linewidth}
\includegraphics[width=\linewidth]{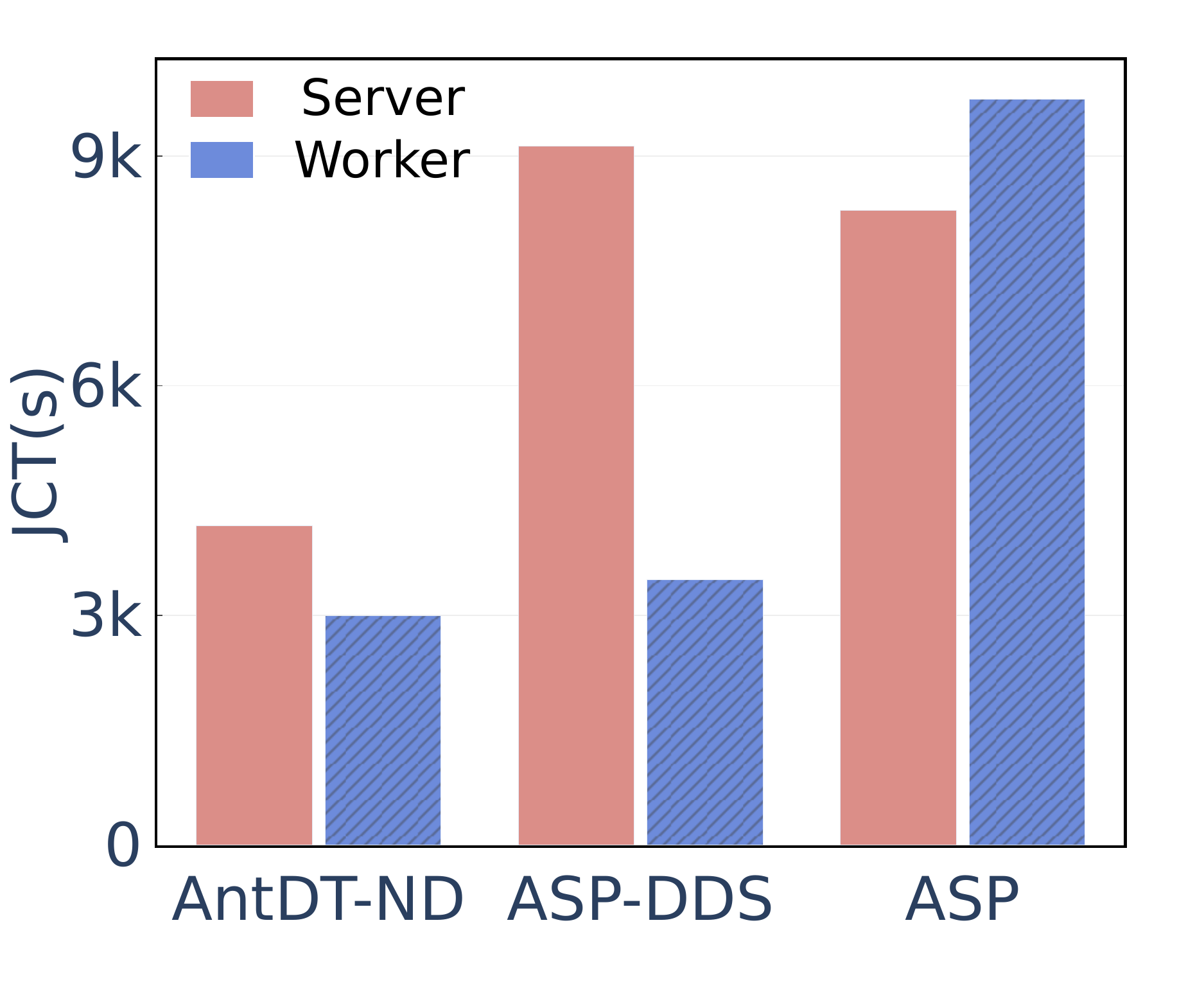}
\caption{JCT for methods in ASP training. The black bar is JCT in worker stragglers and the red bar is JCT in server stragglers.}
\label{fig:jct-asyn-comparision}
\end{minipage}%
\end{figure}

\begin{figure}
\begin{minipage}[t]{0.45\linewidth}
\includegraphics[width=\linewidth]{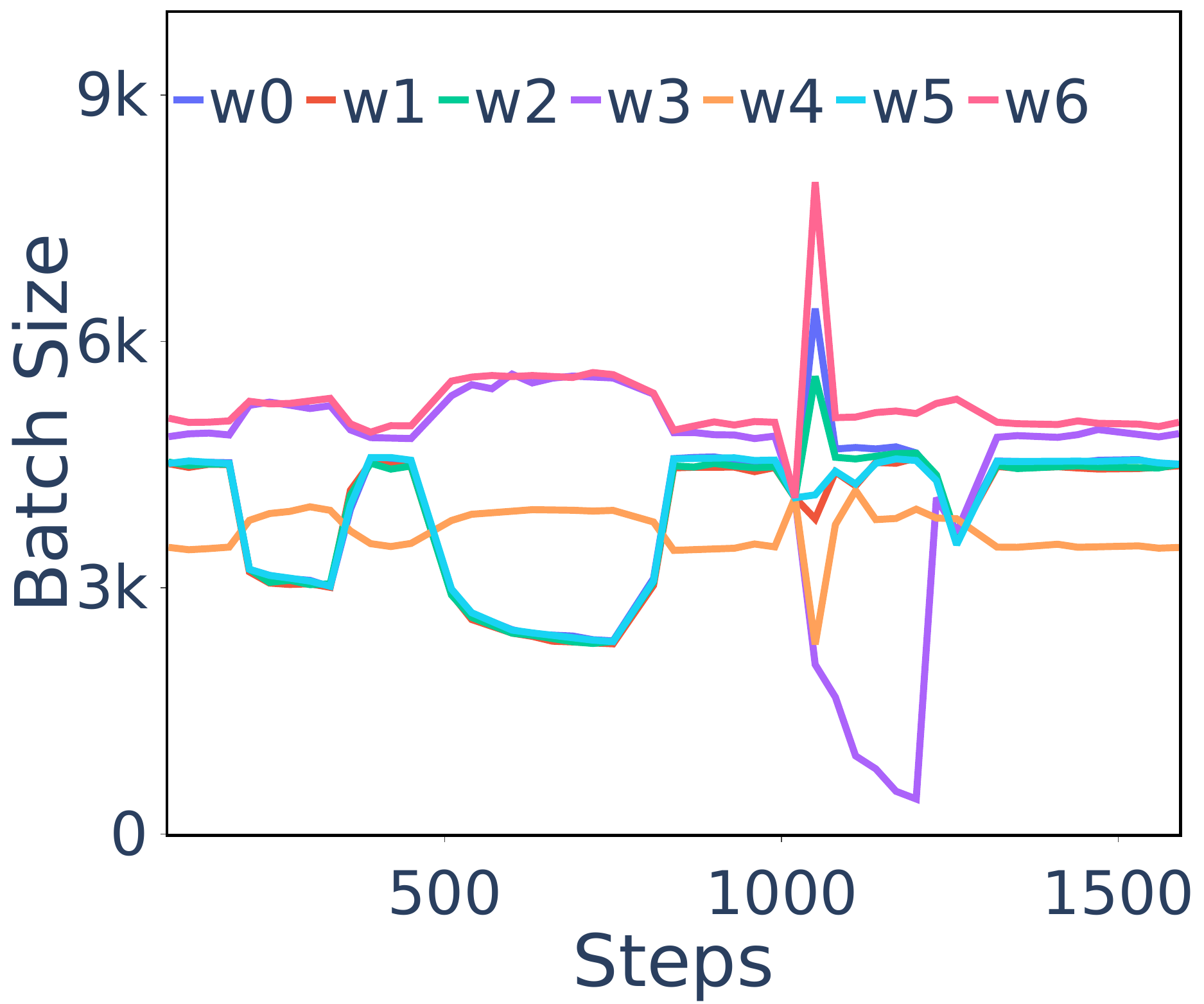}
\caption{Batch size adjustment among workers using AntDT-ND.}
\label{fig:batch-size-adjustment}
\end{minipage}
\hfill
\begin{minipage}[t]{0.45\linewidth}
\includegraphics[width=\linewidth]{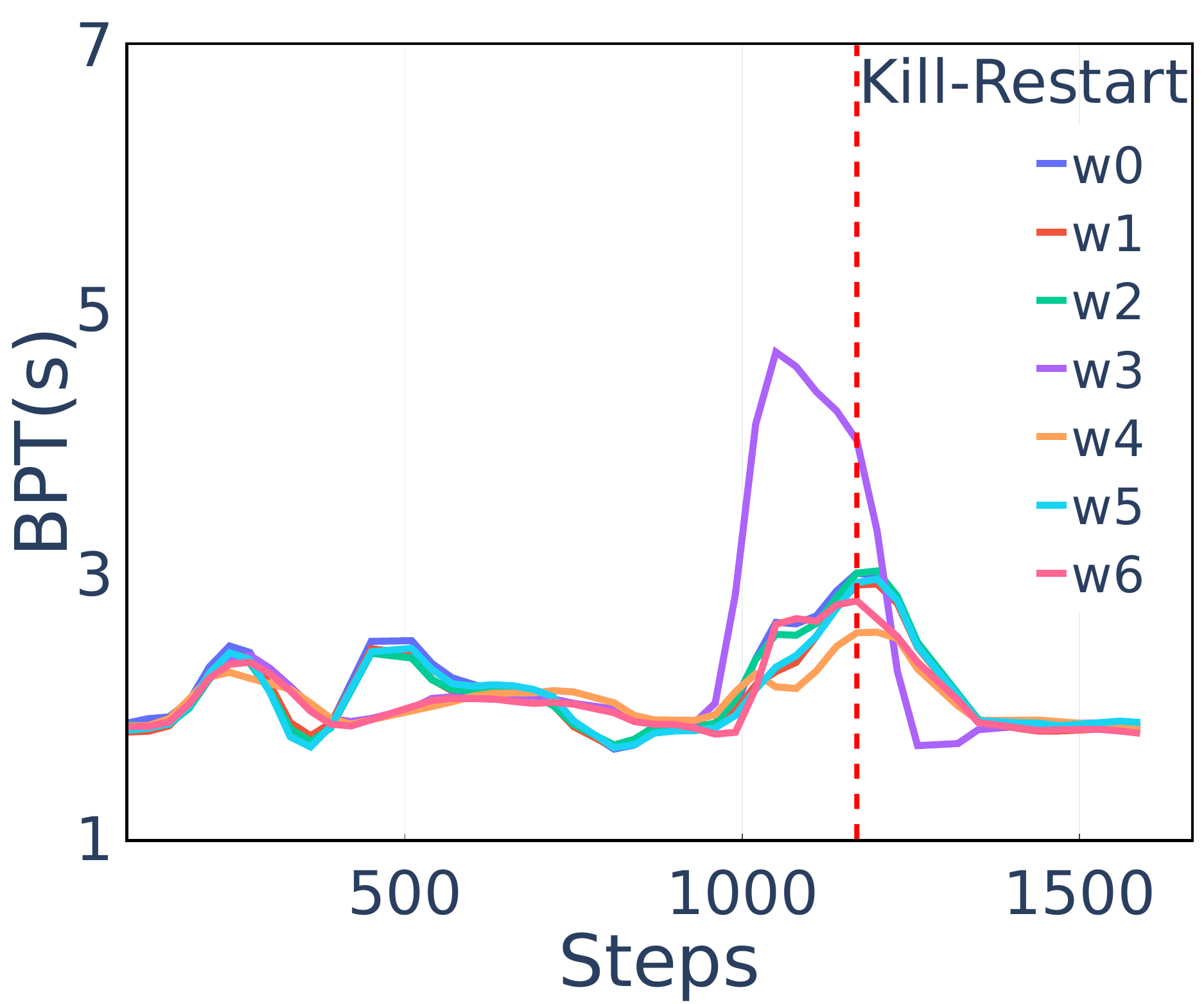}
\caption{BPT of workers using AntDT-ND.}
\label{fig:bpt-workers-kill-restart}
\end{minipage}%
\end{figure}

\subsubsection{Training Efficiency}
We first assess the training efficiency of AntDT-ND in BSP and ASP training in either worker or server stragglers scenarios. We insert the transient and persistent stragglers into the worker threads by setting $SleepDuration$ to 1.5 seconds and $Intensity$ to 0.8. 

\paragraph{Worker Stragglers}
As shown in Fig. \ref{fig:jct-syn-comparision}, in the BSP training, the native BSP is significantly hindered by the worker stragglers and spends 8144 secs (more than 2 hours) in training. The AntDT-ND in BSP training could shorten the overall training time by 24\%, 44\%, and 51\% on average compared with Backup Workers, LB-BSP, and BSP. Fig. \ref{fig:batch-size-adjustment} and \ref{fig:bpt-workers-kill-restart} further illustrate that the dynamical batch size adjustment could increase the batch size of fast workers and decrease the slow workers' to level up all the workers' batch processing time (only part of nodes are selected for visualization). However, the slowest worker, worker 3 or w3 (the purple line in Fig. \ref{fig:batch-size-adjustment} and \ref{fig:bpt-workers-kill-restart}), still has a significant performance gap in BPT against other workers even though we reduce the batch size to a relatively small size, which calls for KILL\_RESTART. We notice that the BPT of other workers is pretty high over the period because all other workers have to undertake more samples given the fixed global batch size, resulting in low global throughput. In contrast to dynamically adjusting the batch size, Backup Worker neglects the stale gradients from the straggling workers, and AntDT-ND additionally reboots the worker 3. Thus, the BPT of worker 3 is soon pulled back to the same level as other workers. This explains why LB-BSP which only uses batch size adjustment does not work as well as AntDT-ND.

In the ASP training, AntDT-ND outperforms the ASP-DDS and ASP by 16\% and 325\%, shown in Fig. \ref{fig:jct-asyn-comparision}. The primary contribution comes from the \emph{DDS service} to adaptively adjust the workloads among the workers via dynamic data shards. The secondary contribution is the \textit{KILL\_RESTART} action, which makes effects by replacing the persistent straggler. 

\begin{table}[t]
\fontsize{6.5}{9}\selectfont
\caption{JCT(seconds) under AntDT-ND and BSP when varying straggler intensity (SI) on the worker or server side.}
\resizebox{0.48\textwidth}{!}{\begin{tabular}{|l|lll|lll|}
\hline
\multicolumn{1}{|c|}{\textbf{}} & \multicolumn{3}{c|}{\textbf{Worker Stragglers}}                      & \multicolumn{3}{c|}{\textbf{Server Stragglers}}                     \\ \hline
\multicolumn{1}{|c|}{\textbf{SI}} & \multicolumn{1}{c}{BSP} & \multicolumn{1}{c}{AntDT-ND} & Speedup & \multicolumn{1}{c}{BSP} & \multicolumn{1}{c}{AntDT-ND} & Speedup \\ \hline
\textbf{0.1}                    & 4312s$\pm$34s  & 3745s$\pm$26s & +10.3\%  & 4627$\pm$51s   & 3636$\pm$15s & +27.3\%  \\ \hline
\textbf{0.3}                    & 4832s$\pm$45s  & 3790s$\pm$40s & +27.5\%  & 6087s$\pm$79s  & 3863$\pm$23s & +57.6\%  \\ \hline
\textbf{0.5}                    & 6004s$\pm$90s  & 3859s$\pm$36s & +55.6\%  & 7365s$\pm$125s & 3944$\pm$24s & +84.4\%  \\ \hline
\textbf{0.8}                    & 8144s$\pm$147s & 3982s$\pm$43s & +104.5\% & 8440s$\pm$194s & 4064$\pm$45s & +107.6\% \\ \hline
\end{tabular}}
\label{table:varying_density}
\end{table}

\begin{figure}
\begin{minipage}[t]{0.45\linewidth}
\includegraphics[width=\linewidth]{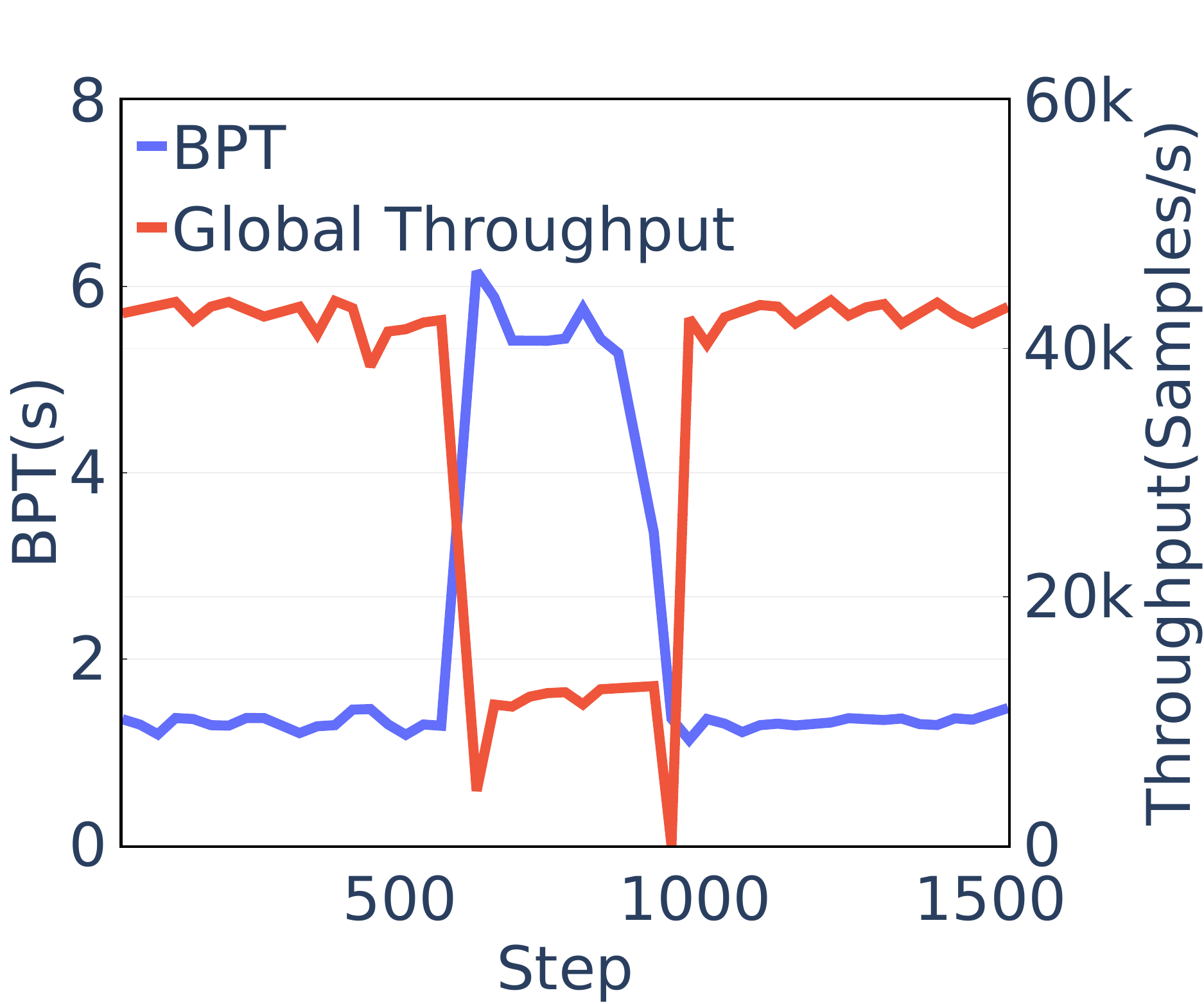}
\caption{BPT of the slow server and global throughput using AntDT-ND in BSP training.}
\label{fig:bpt-servers-kill-restart}
\end{minipage}%
\hfill
\begin{minipage}[t]{0.24\textwidth}
  \includegraphics[width=\linewidth]{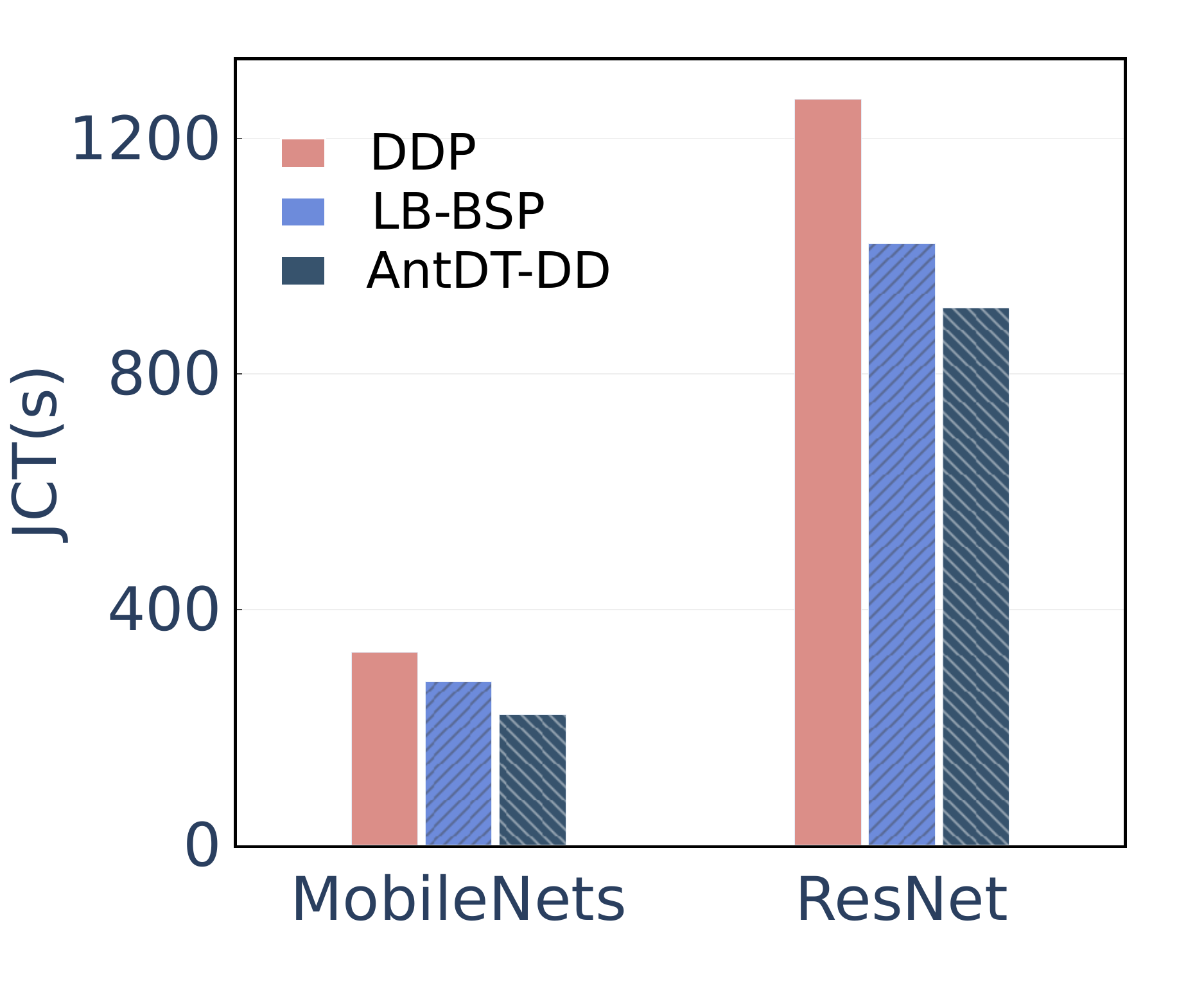}
  \caption{JCT of DDP, LB-BSP, and AntDT-DD in heterogeneous GPU cluster.}
  \label{fig:gpu_jct}
\end{minipage}%
\end{figure}

\begin{figure*}
\begin{minipage}[t]{0.24\linewidth}
\includegraphics[width=\linewidth]{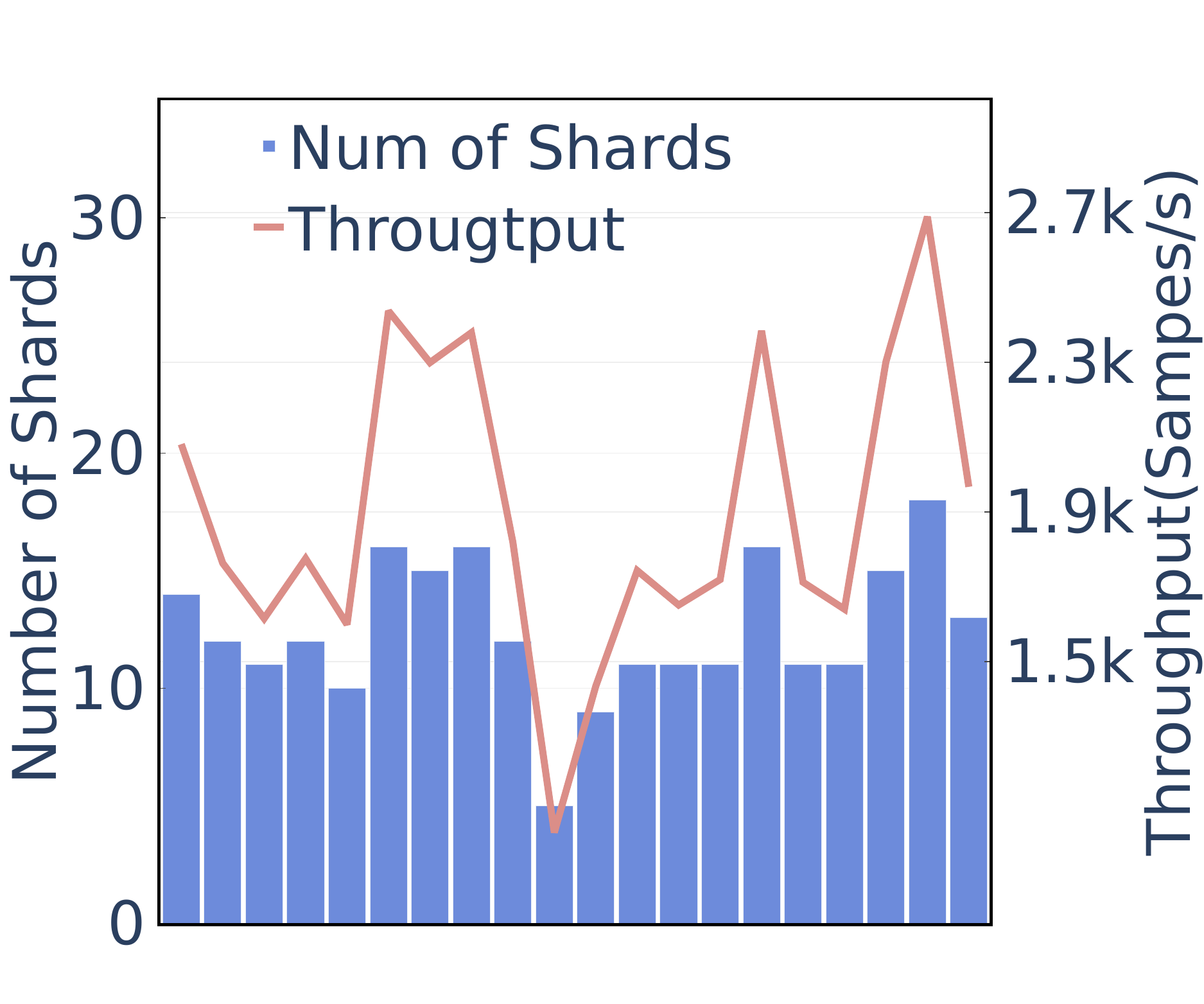}
\caption{The number of data shards against the workers' throughput in ASP-DDS}
\label{fig:data-shards-asp-dds}
\end{minipage}%
\hfill
\begin{minipage}[t]{0.24\linewidth}
\includegraphics[width=\linewidth]{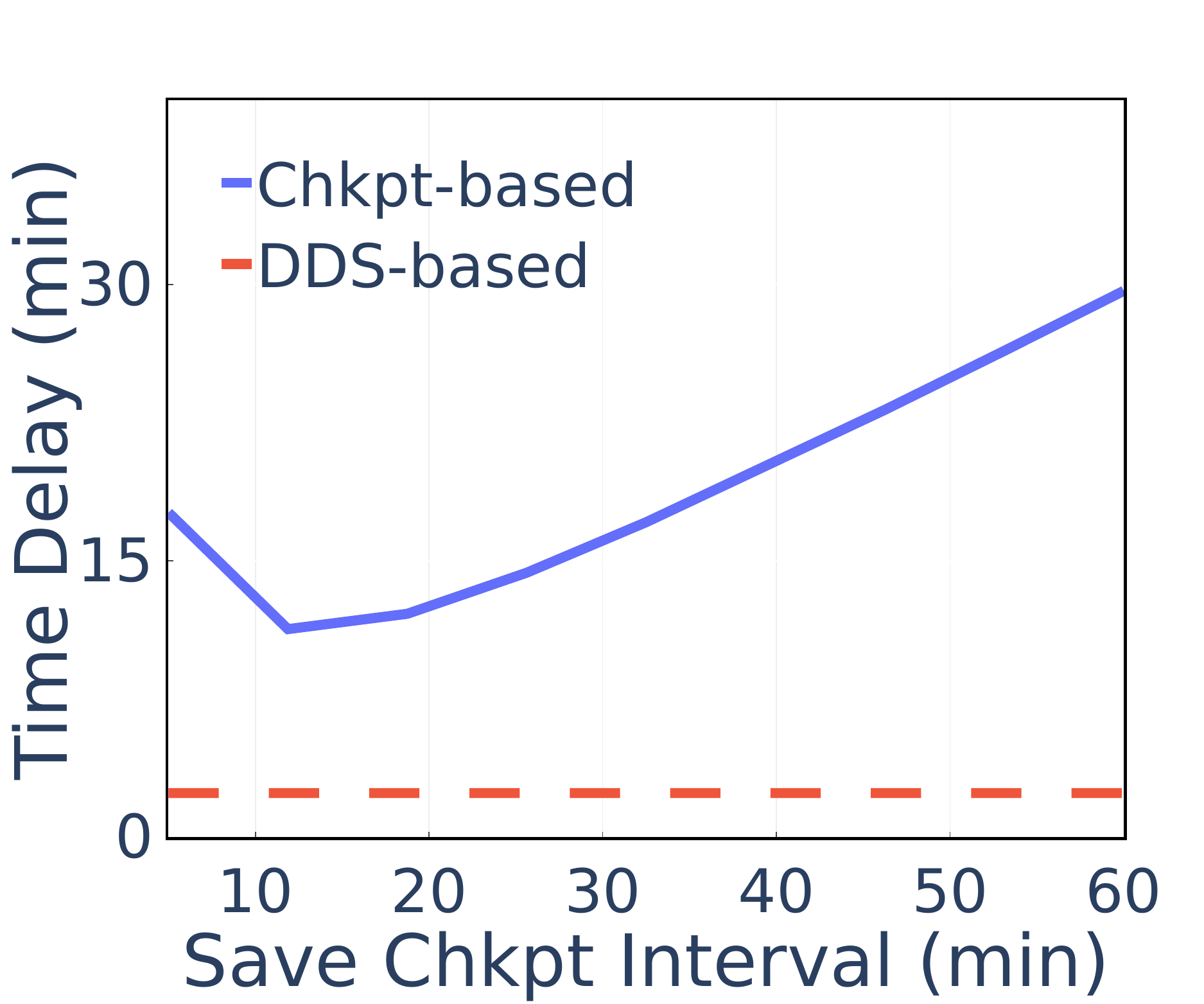}
\caption{Time Delay between Checkpoint-based and DDS-based methods in worker KILL\_RESTART.}
\label{fig:time-delay-chkpt}
\end{minipage}
\hfill
\begin{minipage}[t]{0.24\textwidth}
  \includegraphics[width=\linewidth]{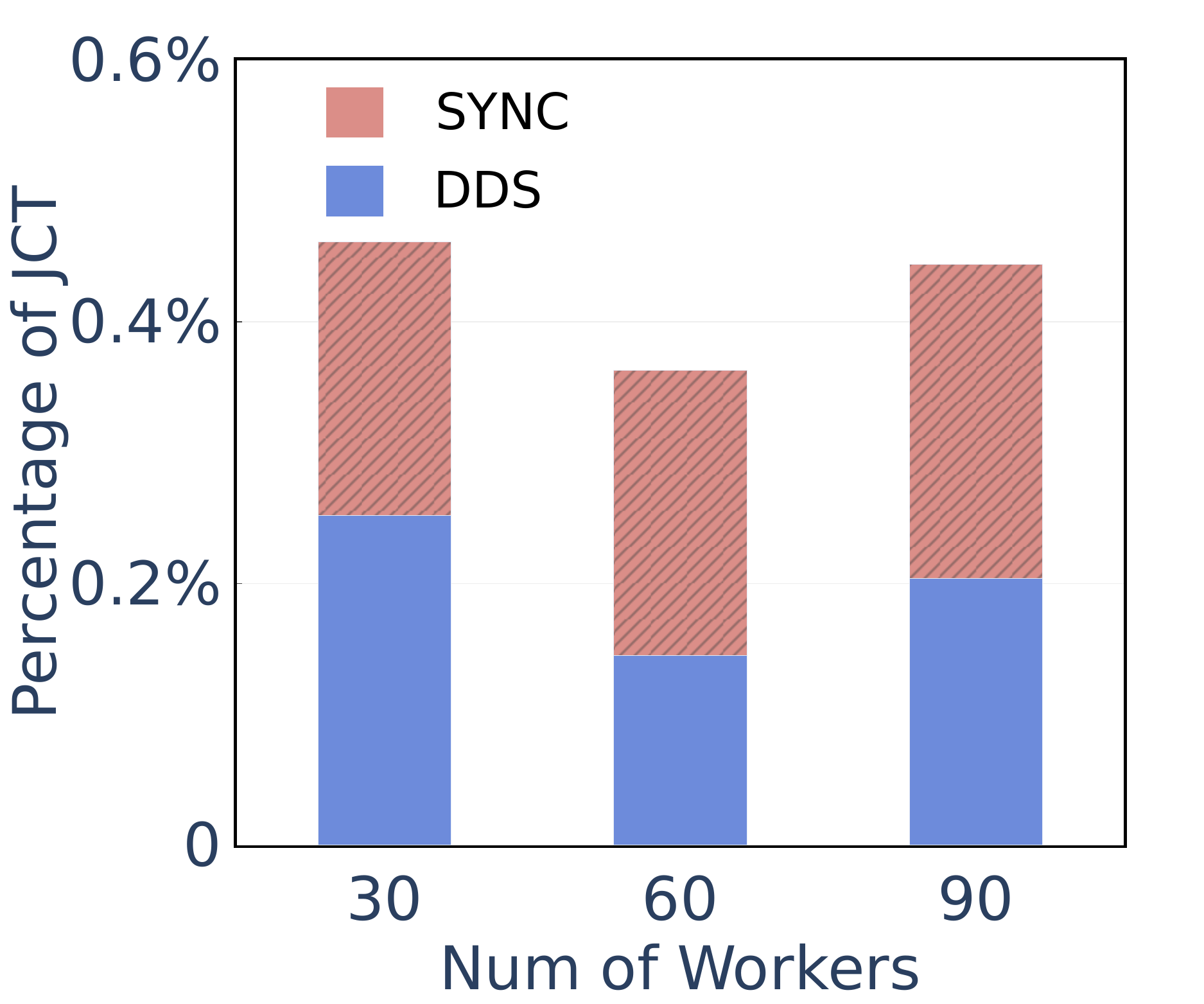}
  \caption{The overhead of AntDT. The blue bar is \emph{DDS}'s and the red bar is synchronization mechanism's.}
  \label{fig:overhead-antdt-growing-size-cpu}
\end{minipage}%
\hfill
\begin{minipage}[t]{0.24\textwidth}
  \includegraphics[width=\linewidth]{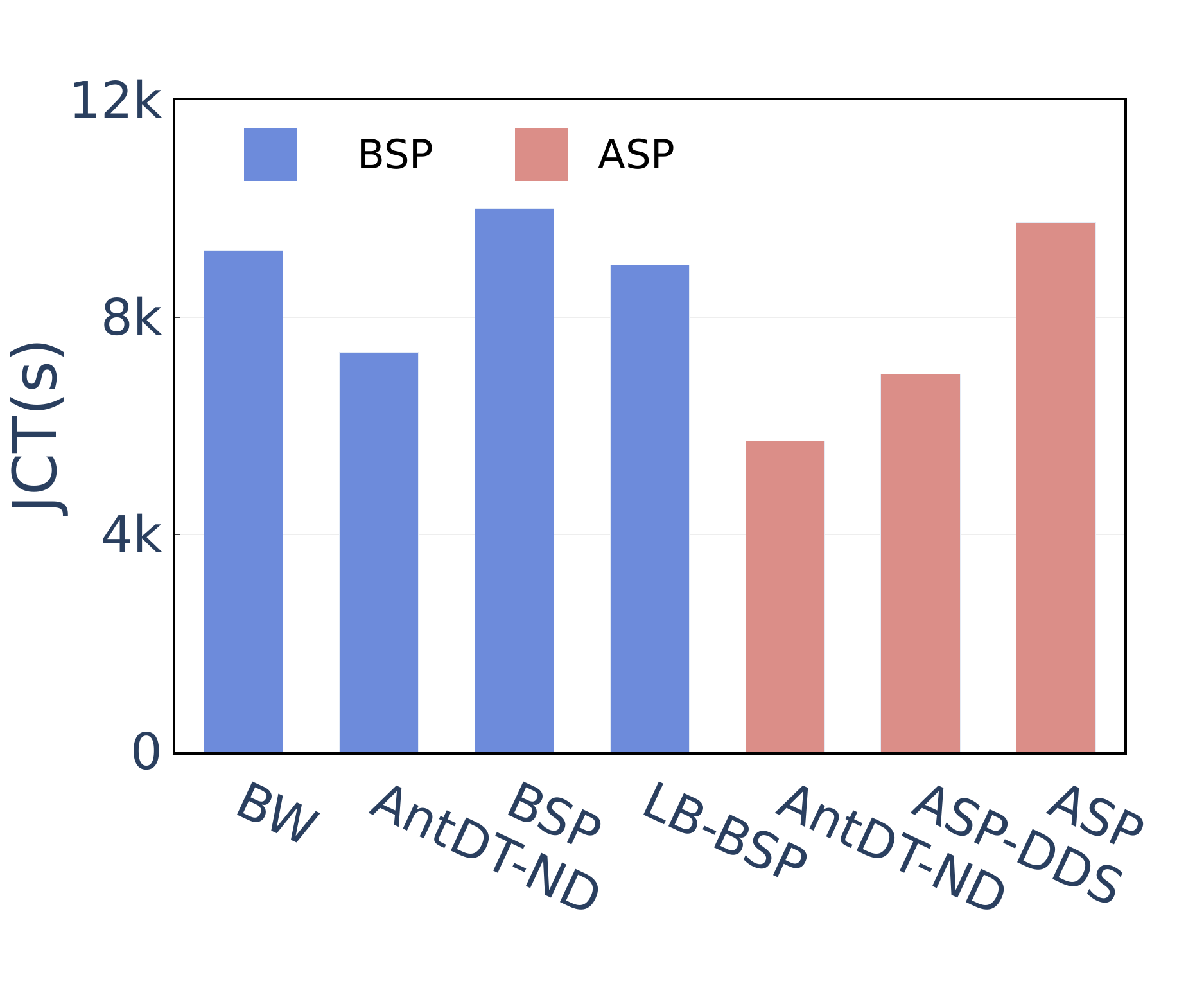}
  \caption{JCT in the production cluster. The blue bar is BSP training and the red is ASP training.}
  \label{fig:jct-sync-training-non-dedicated-cpu}
\end{minipage}%
\end{figure*}

\paragraph{Server Stragglers} 
Compared with injecting the sleep processes in multiple workers, the experiment injects it into one server node since only one straggling server could barrier the whole training phase, and we only insert the persistent straggler into the server node.

As depicted in Fig. \ref{fig:jct-syn-comparision}, AntDT-ND is more than two times faster than all other methods in BSP training. \textcolor{black}{Specifically, AntDT-ND reduces the JCT of LB-BSP, Backup Workers, and BSP by 51\%, 66\%, and 61\% respectively.} In this case, neither LB-BSP nor Backup Worker could alleviate the server straggler problem because one server has to handle the local gradients from all the workers regardless of the change of workloads. Moreover, the Backup Workers method additionally abandons the workers' gradients and worsens the JCT. Fig. \ref{fig:bpt-servers-kill-restart} further illustrates that the BPT of the slow server recovers to normal when the persistent straggler is terminated and restarted. Consequently, the job's global throughput (samples/sec) rebounds after the \textit{KILL\_RESTART} action. In ASP training, we see a similar tendency as shown in Fig. \ref{fig:jct-asyn-comparision}, and the AntDT-ND is two times as fast as the native ASP. Also, we notice that ASP even consumes more time than BSP when there are only server stragglers, which is counterintuitive. This is because ASP requires a higher frequency to update the model parameters on the server side, where any worker will communicate with servers after it completes the local computation.

\subsubsection{Robustness}
We further assess AntDT-ND's robustness regarding training efficiency by analyzing how it performs in increasing straggler intensities. Firstly, we raise the intensity of stragglers from 0.1 to 0.8 on the worker side. Table \ref{table:varying_density} shows that the JCT of native BSP climbs up while there is only a minor increase of the JCT in AntDT-ND. In contrast to BSP, the AntDT-ND speeds up by 10.3\% to 104.5\% when we lift the straggler intensity from 0.1 to 0.8 on the worker side. Additionally, the variation of JCT in AntDT-ND is relatively less than in BSP. Secondly, we exhibit their performance under different straggler intensities on the server node. The JCT of AntDT-ND grows slightly with the increase of straggler intensity while the native BSP shows a sharp increase.

In summary, the AntDT-ND could achieve more than $3\times$ speedup over the existing straggler mitigation methods in either worker or server stragglers scenarios. Additionally, the robustness experiments demonstrate that the AntDT-ND is more stable and robust to the stragglers.

\subsection{Evaluation of AntDT-DD (for Q2)} \label{sec:exp_q2}
We train the model in a mixed series of GPU devices to evaluate how AntDT-DD works in the dedicated cluster with heterogeneous hardware. The synthetic straggler patterns are not inserted into the cluster since V100 and P100 have a natural performance gap. As illustrated in Fig. \ref{fig:gpu_jct}, DDP takes about 1266 seconds, and the LB-BSP takes about 1020 seconds, while AntDT-DD only costs 912 seconds to complete one epoch training of ImageNet using ResNet-101. Thus, AntDT-DD surpasses DDP and LB-BSP by 38.8\%  and 12\% in training speed, given the same amount of training data. For the communication-intensive model like MobileNets, the performance gap enlarges. AntDT-DD runs 25\% and 48.5\% times faster than LB-BSP and DDP in MobileNets. In conclusion, AntDT-DD could achieve up to near $1.5\times$ speedup compared with other methods in the heterogeneous GPU cluster.

\subsection{Evaluation of AntDT framework (for Q3)} \label{sec:exp_q3}
This subsection demonstrates the effectiveness of the AntDT framework in adaptive data allocation,  data integrity, and low time costs compared to the checkpoint-based method during worker failover while maintaining statistical performance.

\subsubsection{Agility of Data Assignment}
Without loss of generality, we take the data shards assignment in the ASP-DDS as an example. Other methods have already used the \emph{DDS service} for the evaluation in the experiments, which helps validate the extensibility of the framework. As displayed in Fig. \ref{fig:data-shards-asp-dds}, the \emph{DDS service} performs well by distributing the data shards to the workers according to their throughput. The number of data shards estimates the actual data consumption and has a consistent trend as the worker throughput since slow workers naturally request few data shards.

\subsubsection{Data Integrity} 
This subsection evaluates the data integrity of the AntDT. Firstly, we confirm that the data is not lost during the training by counting the data shards consumed even when there are failovers from \textit{KILL\_RESTART} actions and retryable failures. The total number of ``DONE" shards is equal to $\lceil \frac{N}{BM} \rceil$ in each experiment when several KILL\_RESTART actions and failovers occur, which guarantees the data integrity. Secondly, we report that the AUC of the model is 0.794 on the test data, which is consistent with the AUC result when there are no failovers in training XDeepFM on the Crieto dataset.

\subsubsection{Time Cost in Worker KILL\_RESTART}
During the failover of worker KILL\_RESTART, the time delay in the traditional checkpoint-based method consists of two parts (apart from the node initialization and pending time). The first part is the time spent in saving the checkpoints (including data reading states), and users usually save the checkpoints every given time interval. The second part is re-computing the data for all workers from where the last checkpoint was saved. In contrast, the time delay in \emph{DDS}-based KILL\_RESTART mainly comes from re-computing the data shards allocated to the crashed node, and the synchronization duration in \emph{DDS} is negligible. As shown in Fig. \ref{fig:time-delay-chkpt}, the time delay in the \emph{DDS}-based method is around two minutes, and the checkpoint-based method is 17 minutes even when we keep a high saving frequency ( e.g., saving checkpoints every 5 minutes), and it soon drops to the lowest point. After that, the time delay rockets with the longer save checkpoints interval.

\subsection{Overhead Analysis and Scalability (for Q4)} \label{sec:exp_q4}
We further assess the overhead of the AntDT framework when it scales out to hundreds of nodes as illustrated in Fig. \ref{fig:overhead-antdt-growing-size-cpu}. \textcolor{black}{Particularly, the runtime duration required for solving optimization problems is negligible, even at large scales with 1000 workers. These durations typically range in the milliseconds level and do not block the training process.} The overhead of AntDT mainly comes from the \emph{Stateful DDS} service and synchronization mechanism in the \emph{AntDT Agent}. We measure the overhead via the total synchronization time divided by the JCT during the process, regardless of the asynchronous threads. We take the delay time of AntDT-ND in BSP training as an example since the batch size adjustment procedure only runs once in AntDT-DD, and the overhead could be negligible. We separately report the percentage of overhead in three scales of non-dedicated production CPU clusters. In the small cluster-C, the delay time takes account of 0.46\% of JCT, which consists of 55\% state synchronization in \emph{Stateful DDS} and 45\% duration in the synchronization mechanism. In the medium cluster-C and large cluster-C, the percentage of overhead slightly fluctuates and is all less than 0.44\% of the overall JCT. In a nutshell, the experiment results verify that the overhead of the AntDT framework is almost negligible and could scale out in large-scale distributed training.

\subsection{Industrial Deployment}

At Ant Group, we integrate the AntDT into the in-house training framework to support thousands of machine learning training jobs daily across various scenarios, including Advertising, Search, Recommendation, and Risk Control applications. \textcolor{black}{To evaluate the performance of AntDT, we conducted an A/B test on 30\% of the training jobs in a non-dedicated production CPU cluster over three days. Fig. \ref{fig:jct-sync-training-non-dedicated-cpu} illustrates the results for BSP training in the Parameter Server. In terms of the JCT across all jobs, AntDT-ND outperformed BSP, Backup Workers, and LB-BSP methods by 26.5\%, 10\%, and 8\% respectively. In ASP training, AntDT-ND demonstrated an average performance improvement of 41\% over ASP and 17.5\% over ASP-DDS. It is important to note that these jobs consisted of both normal and straggling jobs. We compared the average Job Completion Time across all jobs using different methods. This was necessary because it is not possible to differentiate normal jobs from straggling jobs when employing straggling mitigation techniques. Furthermore, during scheduled production training for our homepage recommendation scenario, we encountered a severe straggler problem. Notably, we observed a remarkable five-fold reduction in the Job Completion Time for training our ranking model when utilizing AntDT-ND instead of TensorFlow BSP. This resulted in a decrease from approximately 27.8 hours to just 5.4 hours.  This particular job involved the utilization of 280 CPU nodes, consisting of 230 workers and 50 server nodes, and required training on billions of samples.}

\section{Conclusion}
This paper introduces AntDT, a unified framework to systematically address the straggler problems during distributed deep learning training in real-world scenarios. Firstly, we emphasize that the AntDT offers a self-adaptive framework to easily customize the straggler mitigation solutions using different straggler mitigation methods without considering the complicated data allocation and fault tolerance mechanism. Secondly, this work proposes two practical solutions using our framework as running examples to resolve various types of stragglers in actual production clusters, which outperform other SOTA methods over $3\times$. Our extensive experimental results and industrial deployment clearly echo the effectiveness of the framework in real industrial scenarios.

\bibliographystyle{IEEEtran}
\bibliography{ref}

\end{document}